\documentclass[pre,longbibliography,onecolumn,preprintnumbers,notitlepage,amsmath,amssymb,groupedaddress]{revtex4-2}

\usepackage{amsmath}
\usepackage{amssymb}
\usepackage{amsbsy}
\usepackage{graphicx}
\usepackage{xcolor}

\usepackage{mathrsfs}

\usepackage{enumerate}
\usepackage{mathtools}

\newcommand{\moy}[1]{\left\langle #1 \right\rangle}

\usepackage{physics}

\usepackage{soul}

\usepackage{dsfont}
\usepackage{bbm}

\newcommand{\dt}[2]{\ensuremath{\frac{\dd #1}{\dd #2}}}

\def\e{e}
\def\I{\mathrm{i}}

\newcommand{\sg}[1]{\ensuremath{\mathrm{sign}\left( #1 \right)}}

\DeclareMathOperator{\erfc}{erfc}

\definecolor{darkblue}{rgb}{0,0,0.6}
\definecolor{darkred}{rgb}{0.6,0,0}
\usepackage[colorlinks=true,urlcolor=darkblue,citecolor=darkblue,linkcolor=darkred]{hyperref}

\def\rb{\bar\rho}
\def\nuv{\vec{\nu}}
\def\muv{\vec{\mu}}
\def\rv{\vec{r}}

\usepackage{bm}
\usepackage[capitalize]{cleveref}

\newcommand{\cor}[1]{\mathcal{#1}}
\newcommand{\dslash}[1]{\frac{\mathrm{d}^D q}{(2\pi)^D}}
\newcommand{\T}[1]{\text{#1}}
\newcommand{\n}{\nonumber}
\def \rpn {_{\vec r+\vec \nu}}
\def \rmn {_{\vec r-\vec \nu}}
\def \r {_{\vec r}}
\def \s {_{\vec s}}
\def \z {^{(0)}}
\def \o {^{(1)}}
\def \t {^{(2)}}
\def \q {_{\vec q}}
\newcommand{\rev}[1]{#1}

\begin{document}

\title{Current fluctuations in the symmetric exclusion process\texorpdfstring{\\}{} beyond the one-dimensional geometry}

\author{Th\'eotim Berlioz}
\affiliation{Sorbonne Universit\'e, CNRS, Laboratoire de Physique Th\'eorique de la Mati\`ere Condens\'ee (LPTMC), 4 Place Jussieu, 75005 Paris, France}

\author{Davide Venturelli}
\affiliation{Sorbonne Universit\'e, CNRS, Laboratoire de Physique Th\'eorique de la Mati\`ere Condens\'ee (LPTMC), 4 Place Jussieu, 75005 Paris, France}

\author{Aur\'elien Grabsch}
\affiliation{Sorbonne Universit\'e, CNRS, Laboratoire de Physique Th\'eorique de la Mati\`ere Condens\'ee (LPTMC), 4 Place Jussieu, 75005 Paris, France}

\author{Olivier B\'enichou}
\affiliation{Sorbonne Universit\'e, CNRS, Laboratoire de Physique Th\'eorique de la Mati\`ere Condens\'ee (LPTMC), 4 Place Jussieu, 75005 Paris, France}

\begin{abstract}
    The symmetric simple exclusion process (SEP) is a paradigmatic model of transport, both in and out-of- equilibrium. In this model, the study of currents and their fluctuations has attracted a lot of attention. In finite systems of arbitrary dimension, both the integrated current through a bond (or a fixed surface), and its fluctuations, grow linearly with time. Conversely, for infinite systems, the integrated current displays different behaviours with time, depending on the geometry of the lattice. For instance, in 1D the integrated current fluctuations are known to grow sublinearly with time, as $\sqrt{t}$. Here we study the fluctuations of the integrated current through a given lattice bond beyond the 1D case by considering a SEP on higher-dimensional lattices, and on a comb lattice which describes an intermediate situation between 1D and 2D. We show that the different behaviours of the current originate from qualitatively and quantitatively different current-density correlations in the systems, which we compute explicitly.
\end{abstract}

\maketitle

\tableofcontents

\newpage

\section{Introduction}

The characterisation of currents and their fluctuations is a key problem in statistical physics, which has been the focus on many works over the last decades. In this context, the analysis of simple microscopic models plays a central role, as it allows to obtain explicit results~\cite{Derrida:1998,Derrida:1998a,Bodineau:2004,Derrida:2009,Derrida:2009a}, which can serve as a basis for the development of more general macroscopic descriptions, such as the macroscopic fluctuation theory (MFT)~\cite{Bertini:2015}.

Among such microscopic models, the symmetric exclusion process (SEP) has achieved a paradigmatic status~\cite{Chou:2011,Mallick:2015}, and has received considerable attention since its introduction by Spitzer in 1970~\cite{Spitzer:1970}. It is a model of particles on a lattice, which randomly hop with unit rate to one of the neighbouring sites, only if the site is empty, to mimick hard-core interaction. This model has been the object of recent and important advances, especially in the infinite one-dimensional geometry~\cite{Imamura:2017,Imamura:2021,Poncet:2021,Grabsch:2022,Grabsch:2023,Mallick:2022,Mallick:2024}. In this geometry, the full distribution of the integrated current $Q_t$ through a given bond (corresponding to the total number of particles that have crossed this bond in a given direction, minus the number in the other direction, up to time $t$) was computed in~\cite{Derrida:2009}. In particular, it was shown that at long times all the cumulants of $Q_t$ grow as $\sqrt{t}$. This scaling of the cumulants originates from the fact that the correlations between the current $Q_t$ and the density of particles in the system $\rho(x,t)$ are not stationary. These correlations have recently been computed explicitly in the long-time limit~\cite{Grabsch:2022,Grabsch:2023,Mallick:2022,Mallick:2024}.

In higher dimensions or for other lattices, most studies were conducted on finite systems connected to reservoirs, such as in Refs.~\cite{Bodineau:2008,Akkermans:2013}. In that case, the fluctuations of the integrated current always grow linearly with time, regardless of the dimension or the geometry of the lattice. To the best of our knowledge, no results are available on the impact of the geometry and dimension on current fluctuations in infinite systems.

Here we consider the fluctuations of the integrated current $Q_t$ in the SEP through a given bond of an infinite lattice, initially at equilibrium at density $\rb$. We focus on two examples which illustrate different constraints on the dynamics of the particles: (i) the comb lattice, first introduced to represent diffusion in critical percolation clusters~\cite{Weiss:1986}, in which the particles can bypass each other (unlike in 1D), but cannot loop around a bond, so that each particle can at most contribute to $Q_t$ by $\pm 1$; (ii) lattices of dimensions 2 and higher which contain loops, so that a given particle can give an arbitrarily high contribution to $Q_t$. 
We first show how these different dynamical constraints lead to different behaviours of the correlations of $Q_t$ with the density of surrounding particles. In turn, these correlations lead to different behaviours of the current fluctuations, which we quantify explicitly.

The article is organised as follows. We first recall the known results in the one-dimensional case in Section~\ref{sec:1D}. We then turn to the comb lattice in Section~\ref{sec:Comb}, and provide both a microscopic and a macroscopic derivation of the current fluctuations
--- the
macroscopic approach presenting the advantage of being applicable to other models and higher-order cumulants of $Q_t$. In Section~\ref{sec:2D} we finally consider lattices in two dimensions and more, using two different microscopic approaches giving access to different correlation functions. The physical meaning of these different correlations is discussed in Section~\ref{sec:DiscCurrProf}. We finish with some concluding remarks in Section~\ref{sec:Conclusion}.

\section{Known results in the 1D case}
\label{sec:1D}

We first review a few known results about the fluctuations of $Q_t$ through the bond linking sites $0$ and $1$ (denoted $0-1$ from now on) for a SEP in a one-dimensional infinite lattice, initially at equilibrium at density $\rb$. 
By symmetry, $\moy{Q_t}=0$, and the fluctuations of the current read~\cite{Ferrari:1994,Derrida:2009}
\begin{equation}
    \label{eq:FlucCurrent1D}
    \moy{Q_t^2}_{\mathrm{1D}}
    \underset{t \to \infty}{\simeq}
    \frac{2\rb(1-\rb)}{\sqrt{\pi}} 
    \sqrt{t}
    \:.
\end{equation}
The current fluctuations are proportional to $\rb(1-\rb)$, which is expected due to the particle/hole symmetry of the SEP~\cite{Derrida:2004}, and the fact that the current of particles is the opposite of the current of vacancies. The form~\eqref{eq:FlucCurrent1D} indicates that the current vanishes both at low density (no particles) and at high density (all sites are occupied, so no motion is possible).

\begin{figure}
    \centering
    \raisebox{0.12\textwidth}{\includegraphics[width=0.4\textwidth]{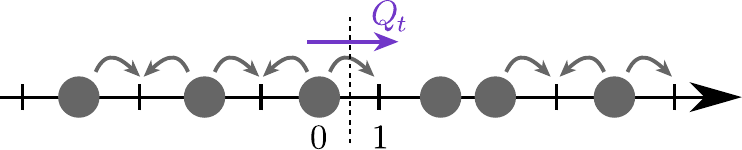}}
    \hspace{0.05\textwidth}
    \includegraphics[height=0.3\textwidth]{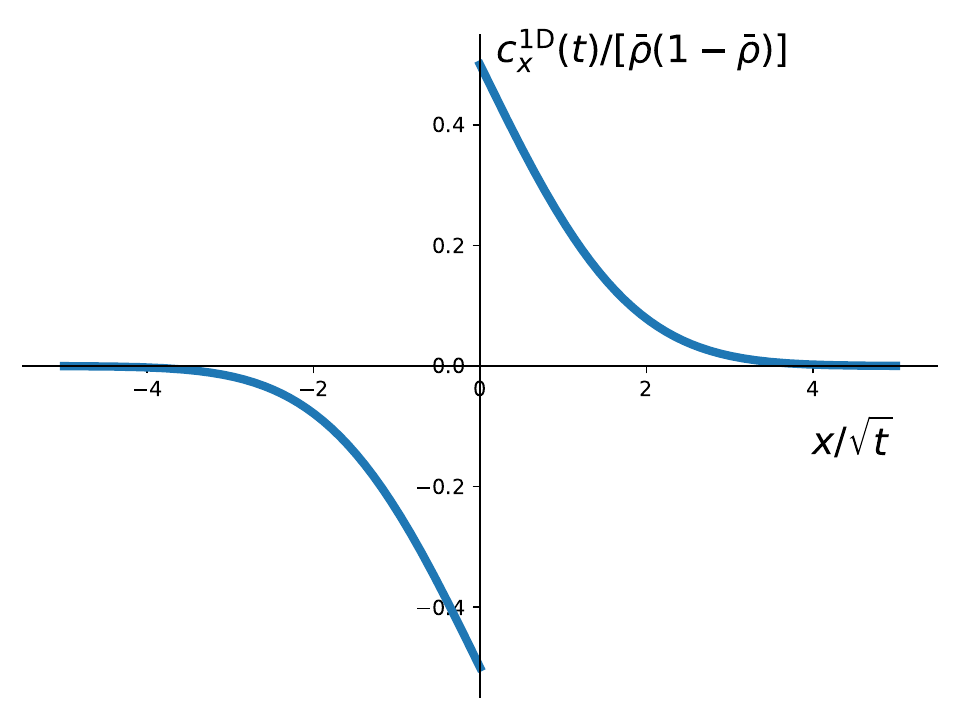}
    \caption{Left: SEP on a $1$D infinite lattice. The particles can jump to an empty neighbouring site with unit rate in each direction. The integrated current $Q_t$ is measured on the bond between sites $0$ and $1$. Right: correlation profile $c_x^{\mathrm{1D}}(t)$ at long times, as a function of the scaling variable $x/\sqrt{t}$, given by~\eqref{eq:Correl1D}.}
    \label{fig:1D}
\end{figure}

Importantly, the $\sqrt{t}$ behaviour of the fluctuations~\eqref{eq:FlucCurrent1D} yields much smaller fluctuations at large times than does the linear behaviour in $t$ found for a finite 1D system~\cite{Bodineau:2004}.
The origin of this anomalous scaling can be traced back to the spatial structure of the correlations between the integrated current $Q_t$ and the occupations of the sites $\eta_r(t)$ ($\eta_r(t) = 1$ if site $r$ is occupied at time $t$, $0$ otherwise)~\cite{Poncet:2021,Grabsch:2022},
\begin{equation}
    c_r^{\mathrm{(1D)}}(t) \equiv \mathrm{Cov}(Q_t, \eta_r(t))
    = \moy{Q_t \eta_r(t)} - \moy{Q_t} \moy{\eta_r(t)}
    = \moy{Q_t \eta_r(t)}
    \:.
\end{equation}
These correlations have been computed recently~\cite{Poncet:2021,Grabsch:2022}, and take the following scaling form at long times,
\begin{equation}
    \label{eq:Correl1D}
    c_r^{\mathrm{(1D)}}(t)
    \underset{t \to \infty}{\simeq}
    \left\lbrace
    \begin{array}{cl}
    \displaystyle
        \frac{\rb(1-\rb)}{2} \erfc \left( \frac{r}{2 \sqrt{t}} \right) & \text{for } r \geq 1
        \:, \\[0.4cm]
    \displaystyle
        -\frac{\rb(1-\rb)}{2} \erfc \left( -\frac{r}{2 \sqrt{t}} \right) & \text{for } r \leq 0
    \:,
    \end{array}
    \right.
\end{equation}
with $\erfc$ the complementary error function (see Fig.~\ref{fig:1D}). This result shows that the correlations are never stationary and grow with time on a distance that scales with $\sqrt{t}$. 
This is due to the fact that, in 1D, the integrated current $Q_t$ is strongly correlated with the density of particles around the bond in which the current is measured: a fluctuation that increases $Q_t$ is necessarily due to an increase of particles crossing the bond $0-1$, leading to an accumulation of particles on the positive sites ($c_r(t) \geq 0$), and a depletion on the negative sites ($c_r(t) \leq 0$). Since this imbalance cannot be absorbed by reservoirs or periodic boundary conditions, larger fluctuations require more particles to cross from left to right, pushing more and more particles further on the positive lattice sites due to the exclusion rule.

The fact that the correlations are not stationary in 1D can be at first sight attributed to two factors: (i) the conservation of the initial order of the particles at all times, due to the exclusion rule which mimics hard core interaction; and (ii) the tree-like structure of the 1D lattice, which prevents a given particle from giving a large contribution to $Q_t$. In the following, we investigate the effect of each  of these two contributions by considering different lattices.

\section{The comb lattice}
\label{sec:Comb}

The comb lattice is a 2D infinite lattice in which all the horizontal links have been removed, except on the horizontal axis $y=0$ (see Fig.~\ref{fig:CombQ}, left), called the backbone. The vertical lines connected to the backbone are called the teeth.

Comb structures have been introduced to represent diffusion in critical percolation clusters: the backbone and teeth of the comb mimic the quasi-linear structure and the dead ends of percolation clusters~\cite{Weiss:1986}.
More recently, the comb model has been used to describe transport in real systems such as spiny dendrites~\cite{Mendez:2013}, diffusion of cold atoms~\cite{Sagi:2012}, and diffusion in crowded media~\cite{Hofling:2013}. In the context of interacting particles, exclusion processes on the comb have been used to study the displacement of a tracer particle~\cite{Benichou:2015}. Recently, the statistics of the current of hard core reflecting Brownian particles on a comb, corresponding to the low-density limit of the SEP studied here, has been characterised~\cite{Grabsch:2023a}.

Here we study the integrated current $Q_t$ through the bond $(0,0)-(1,0)$ up to time $t$ for a SEP on the comb lattice, starting initially from an equilibrium measure corresponding to having each site independently occupied with probability $\bar\rho$. Compared to the 1D case, the noncrossing constraint is lifted as particles on the backbone can bypass a particle that jumps into a tooth. However, the comb lattice is still a tree, so a given particle cannot cross several times the bond $(0,0)-(1,0)$ consecutively in the same direction.

In the following, we first provide a microscopic derivation of the current fluctuations and the correlation profiles. We then rederive these results within a purely macroscopic description, which could in principle be used to derive higher-order cumulants of $Q_t$, and also applied to other models than the SEP.

\subsection{Microscopic calculation}

\label{comb_microscopic}

Our starting point is the following master equation for the SEP on the comb, describing the evolution of the probability $P_t(\underline{\eta})$ to observe a given configuration $\underline{\eta}=\{\eta_{x,y}\}_{x,y \in \mathbb{Z}^{2}}$ at time $t$: 
\begin{equation}
\label{eq:MasterEqComb}
    \partial_t P_t(\underline{\eta})
    = \sum_{x, y} \left[P(\underline{\eta}^{x,y+}, t) - P(\underline{\eta}, t) \right] + \sum_{x} \left[P(\underline{\eta}^{x+,0}, t) - P(\underline{\eta}, t) \right]  ,
\end{equation}
where we have denoted by $\underline{\eta}^{x,y+}$ the configuration in which the occupations of sites $(x,y)$ and $(x,y+1)$ have been exchanged, and by $\underline{\eta}^{x+,y}$ the configuration in which the occupations of sites $(x,y)$ and $(x+1,y)$ have been exchanged. Note that the equilibrium distribution reads
\begin{equation}
    P_{\infty}(\underline{\eta}) =\underset{x,y}{\prod}p_{\rb}(\eta_{x,y}),
    \label{equilibrium_distribution}
\end{equation}
which factorizes over the sites of $\mathbb{Z}^{2}$, and where $p_{\rb}$ is the Bernoulli measure
\begin{equation}
\left\{
    \begin{array}{ll}
        p_{\rb}(\eta = 1) = \rb, \\
        p_{\rb}\small{(\eta = 0)} = 1-\rb,
    \end{array}
\right.
\end{equation}
in which each site 
is occupied with probability $\rb$ and empty with probability $(1-\rb)$.
The fluctuations of the integrated current $Q_t$ are described by the cumulant generating function
\begin{equation}
    \psi(\lambda, t) = \ln{\left(\moy{e^{\lambda Q_t}}\right)} = \sum_{n=1}^{+\infty}\frac{\moy{Q_t^{n}
    }_c}{n!}\lambda^{n}, 
    \label{generating_function_comb}
\end{equation}
\rev{where we indicated as $\expval{\bullet}_c$ the connected correlation function, and}
where all odd cumulants are null due to the initial equilibrium 
\rev{measure}
and the symmetries of the problem. 
Correlations between the integrated current and bath particles are encoded in the joint cumulant generating function of $(Q_t, \eta_{x,y})$ for any pair $(x, y) \in \mathbb{Z}^{2}$, namely $\ln{\left(\moy{e^{\lambda Q_t + \mu \eta_{x,y}}} \right)}$.
Since $\eta_{x,y} =0$ or $1$, this expression simplifies as
\begin{equation}
    \ln{\left(\moy{e^{\lambda Q_t + \mu \eta_{x,y}}} \right)} = \ln{\left(\moy{e^{\lambda Q_t}}\right)}+\ln{ \left(1+(e^{\mu}-1)\frac{\moy{\eta_{x,y}e^{\lambda Q_t}}}{\moy{e^{\lambda Q_t}}}\right)}.
\end{equation} 
\rev{From this relation, we see that all the correlations between the occupations and the currents are encoded in the correlation functions
\begin{equation}
    \moy{\eta_{x,y}Q_{t}^n}_c 
    \equiv 
    \left. \partial_\mu \partial_\lambda^n
    \ln{\left(\moy{e^{\lambda Q_t + \mu \eta_{x,y}}} \right)}
    \right|_{\mu=0,\lambda=0}
    = \left.\partial_\lambda^n
    \frac{\moy{\eta_{x,y}e^{\lambda Q_t}}}{\moy{e^{\lambda Q_t}}}
    \right|_{\lambda=0}
    \:.
\end{equation}
}
It is thus convenient to introduce the generalized correlation profile~\cite{Poncet:2021,Grabsch:2022}
\begin{equation}
   w_{x,y}(\lambda, t) = \frac{\moy{\eta_{x,y}e^{\lambda Q_t}}}{\moy{e^{\lambda Q_t}}} = \sum_{n=0}^{+\infty}\moy{\eta_{x,y}Q_t^{n}}_c\frac{\lambda^{n}}{n!}
   \:.
   \label{correlation_comb}
\end{equation}
In particular, at first order in $\lambda$, it gives
\begin{equation}
    \label{eq:c_comb}
    c_{x,y}^{\mathrm{(Comb)}}(t) \equiv \moy{\eta_{x,y}(t) \: Q_t}_c
    = \moy{\eta_{x,y}(t) \: Q_t} - \moy{\eta_{x,y}(t)} \moy{Q_t}
    \:.
\end{equation}

The first step to get correlations and fluctuations of the current is to write down the evolution equations of those quantities using the master equation~\eqref{eq:MasterEqComb}. The derivation of the evolution equation for $\psi$ is straightforward, and results in
\begin{equation}
    \partial_t\psi(\lambda, t) = (e^{\lambda}-1)\frac{\moy{\eta_{0,0}(1-\eta_{1,0})e^{\lambda Q_t}}}{\moy{e^{\lambda Q_t}}} + (e^{-\lambda}-1)\frac{\moy{\eta_{1,0}(1-\eta_{0,0})e^{\lambda Q_t}}}{\moy{e^{\lambda Q_t}}}.
     \label{eq:cumulant-eq}
\end{equation}
Let us then consider the evolution equation of $w_{x,y}$.
Particles in the bulk can jump without affecting the current, whereas the current increases by 1 if a particle located at site $(0,0)$ jumps to the empty site $(1,0)$, and decreases by 1 if a particle located at site $(1,0)$ jumps to the empty site $(0,0)$. Then, only the configuration $\underline{\eta}^{0+,0}$ can impact the current. Thus,
in the following it is convenient to consider separately the contributions from
sites in the bulk $(x,y)\neq \{(0,0), (1,0)\}$,
and sites at the boundaries. 
Using the master equation~\eqref{eq:MasterEqComb} we get for the bulk
\begin{multline}
    \partial_t w_{x,y} =\sum_{\mu = \pm1}\nabla_{\mu}^yw_{x,y} \quad + \quad \delta_{y,0}\sum_{\mu = \pm1}\nabla_{\mu}^xw_{x,0}   \\ + \quad (e^{\lambda}-1)\left[\frac{\moy{\eta_{x,y}\eta_{ 0,0}(1-\eta_{1,0})e^{\lambda Q_t}}}{\rev{\moy{ e^{\lambda Q_t}}}} - w_{x,y}\frac{\moy{\eta_{0,0}(1-\eta_{1,0})e^{\lambda Q_t}}}{\rev{\moy{e^{\lambda Q_t}}} } \right] \\
    + \quad (e^{-\lambda}-1)\left[\frac{\moy{\eta_{x,y}\eta_{ 1,0}(1-\eta_{0,0})e^{\lambda Q_t}}}{\rev{\moy{e^{\lambda Q_t}}}} - w_{x,y}\frac{\moy{\eta_{1,0}(1-\eta_{0,0})e^{\lambda Q_t}}}{\rev{\moy{e^{\lambda Q_t}}}} \right],
    \label{eq:profiles_bulk}
\end{multline}
and for the boundaries
\begin{align}
     &\partial_tw_{0,0} =e^{-\lambda}\frac{\moy{\eta_{1,0}(1-\eta_{0,0})e^{\lambda Q_t}}}{\moy{e^{\lambda Q_t}}}-\frac{\moy{\eta_{0,0}(1-\eta_{1,0})e^{\lambda Q_t}}}{\moy{e^{\lambda Q_t}}} -w_{0,0}  \partial_t\psi(\lambda, t) +\sum_{\mu=\pm1}\nabla_{\mu}^yw_{0,0} \quad + \quad \nabla_{-1}^xw_{0,0}, \label{masterboundary0_0} \\
     &\partial_tw_{1,0} =e^{\lambda}\frac{\moy{\eta_{0,0}(1-\eta_{1,0})e^{\lambda Q_t}}}{\moy{e^{\lambda Q_t}}}-\frac{\moy{\eta_{1,0}(1-\eta_{0,0})e^{\lambda Q_t}}}{\moy{e^{\lambda Q_t}}} -w_{1,0}  \partial_t\psi(\lambda, t) +\sum_{\mu=\pm1}\nabla_{\mu}^yw_{1,0} \quad + \quad \nabla_{1}^xw_{1,0},
     \label{masterboundary1_0}
\end{align}
 where we have introduced the discrete gradient acting as
\begin{equation}
\nabla_{\mu}^{y}F(x,y)=F(x,y+\mu)-F(x,y),\qquad \nabla_{\mu}^{x}F(x,y)=F(x+\mu,y)-F(x,y), \qquad \text{with } \mu =\pm1,
\end{equation}
on a generic function $F(x,y)$ of $\mathbb{Z}^{2}$.
The boundary equations \eqref{masterboundary0_0}~and~\eqref{masterboundary1_0} can be further rewritten using the evolution equation~\eqref{eq:cumulant-eq} of the generating function $\psi$ as
\begin{align}
     &\partial_tw_{0,0} =\biggl[e^{-\lambda} - (e^{-\lambda}-1)w_{0,0} \biggr]\frac{\partial_t\psi(\lambda, t)}{e^{-\lambda}-1} +\sum_{\mu=\pm1}\nabla_{\mu}^yw_{0,0} \quad + \nabla_{-1}^xw_{0,0}, \label{masterboundary0} \\
     &\partial_tw_{1,0} =\biggl[e^{\lambda} - (e^{\lambda}-1)w_{1,0} \biggr]\frac{\partial_t\psi(\lambda, t)}{e^{\lambda}-1} +\sum_{\mu=\pm1}\nabla_{\mu}^yw_{1,0} \quad + \nabla_{1}^xw_{1,0} .
     \label{masterboundary1}
\end{align}

We note that the bulk equation~\eqref{eq:profiles_bulk} is not closed, since it involves the higher-order correlation functions $\frac{\moy{\eta_{0,0}\eta_{x,y}e^{\lambda Q_t}}}{\moy{e^{\lambda Q_t}}}$ and $\frac{\moy{\eta_{1,0}\eta_{x,y}e^{\lambda Q_t}}}{\moy{e^{\lambda Q_t}}}$. However, we will see that it yields a closed equation for the correlations $c_{x,y}^{\mathrm{(Comb)}}(t)$~\eqref{eq:c_comb}.
Indeed, expanding $w_{x,y}$ up to first order in $\lambda$,
\begin{equation}
    w_{x,y}(\lambda,t)=\rb +\lambda \: c_{x,y}^{\mathrm{(Comb)}}(t) + O(\lambda^{2}),
\end{equation}
and inserting this expansion into the bulk equation~\eqref{eq:profiles_bulk} leads to:
\begin{equation}
    \partial_t c_{x,y}^{\mathrm{(Comb)}}(t) = \sum_{\mu = \pm1}\nabla_{\mu}^y c_{x,y}^{\mathrm{(Comb)}}(t)+ \delta_{y,0}\sum_{\mu = \pm1}\nabla_{\mu}^x c_{x,y}^{\mathrm{(Comb)}}(t)+\moy{\eta_{x,y}\eta_{0,0}}-\rev{\moy{\eta_{x,y}}\moy{\eta_{0,0}}} -\moy{\eta_{x,y}\eta_{1,0}}+\rev{\moy{\eta_{x,y}}\moy{\eta_{1,0}}}.
\end{equation}
The latter simplifies as
\begin{equation}
     \partial_t c_{x,y}^{\mathrm{(Comb)}}(t) = \sum_{\mu = \pm1}\nabla_{\mu}^y c_{x,y}^{\mathrm{(Comb)}}(t)+ \delta_{y,0}\sum_{\mu = \pm1}\nabla_{\mu}^x c_{x,y}^{\mathrm{(Comb)}}(t)
    \label{diff_teeth}
\end{equation}
upon noting that $\moy{\eta_{x,y}\eta_{x',y'}} =\rb^{2}$, as it follows from inspecting
the stationary distribution $P_\infty$ in Eq.~\eqref{equilibrium_distribution}.
Similarly, the boundary equations~\eqref{masterboundary0}~and~\eqref{masterboundary1} give at leading order in $\lambda$
\begin{align}
     &\partial_t c_{0,0}^{\mathrm{(Comb)}}(t) = -\frac{\partial_t\moy{Q_t^{2}}}{2} + \sum_{\mu = \pm1}\nabla_{\mu}^y c_{0,0}^{\mathrm{(Comb)}}(t)+ \nabla_{-1}^x c_{0,0}^{\mathrm{(Comb)}}(t), \label{boundary_0}\\
      &\partial_t c_{1,0}^{\mathrm{(Comb)}}(t) = \frac{\partial_t\moy{Q_t^{2}}}{2}+ \sum_{\mu = \pm1}\nabla_{\mu}^y c_{1,0}^{\mathrm{(Comb)}}(t)+ \nabla_{1}^x c_{1,0}^{\mathrm{(Comb)}}(t).
      \label{boundary}
\end{align}
Note that, contrary to \cref{diff_teeth} for the bulk, 
which is exact only at equilibrium, the equations verified by $c_{0,0}^{\mathrm{(Comb)}}(t)$ and $c_{1,0}^{\mathrm{(Comb)}}(t)$ remain valid also out of equilibrium --- for instance, if we start initially from a step of density $\rho_+$ for $x > 0$ and $\rho_-$ for $x<0$.

To get the current fluctuations, we expand the evolution equation of the generating function~\eqref{eq:cumulant-eq} up to second order in $\lambda$, finding
\begin{equation}
   \frac{\partial_t\moy{Q_t^{2}}}{2}= c_{0,0}^{\mathrm{(Comb)}}(t)-c_{1,0}^{\mathrm{(Comb)}}(t) +\rb(1-\rb).
   \label{eq:variance}
\end{equation}
Equations~\eqref{diff_teeth}~to~\eqref{eq:variance} now form a closed set which
can be solved exactly by introducing the Laplace transform of the variance 
\begin{equation}
    \kappa(u)=\int_{0}^{+\infty}e^{-ut}\moy{Q_t^{2}}_c \dd t\text{,}
\end{equation}
and the Fourier-Laplace transform of the correlations:
\begin{equation}
\tilde{c}_{k,q}(u)= \sum_{x,y}\int_{0}^{+\infty}e^{-ut}e^{ikx+iqy} c_{x,y}^{\mathrm{(Comb)}}(t)\dd{t}.
\end{equation}
Applying the Laplace transform to \cref{eq:variance} 
we obtain
\begin{equation}
    \frac{u\, \kappa(u)}{2}= \frac{\rb(1-\rb)}{u}+ c_{0,0}^{\mathrm{(Comb)}}(u) - c_{1,0}^{\mathrm{(Comb)}}(u),
    \label{eq:kappa}
\end{equation}
while by Fourier-Laplace transforming the correlation 
equations~\eqref{diff_teeth}~to~\eqref{boundary} we get
\begin{equation}
u\, \tilde{c}_{k,q}(u)=2(\cos{q}-1)\tilde{c}_{k,q}(u)+2(\cos{k}-1)\tilde{c}_{k,y=0}(u)+\frac{\rb(1-\rb)}{u}(e^{ik}-1),
\label{eq:correlation fourier}
\end{equation}
where we have denoted as
\begin{equation}
    \tilde{c}_{k,y=0}(u)= \sum_{x}\int_{0}^{+\infty}e^{-ut}e^{ikx} c_{x,0}^{\mathrm{(Comb)}}(t)\dd{t}
\end{equation}
the Fourier-Laplace transform performed only along the horizontal axis $x$ at $y=0$.
The procedure to obtain an exact expression of $c_{x,y}^{\mathrm{(Comb)}}(t)$ is to derive first the correlations along the backbone ($y=0$), and then to deduce from the bulk equation~\eqref{diff_teeth} the correlations for all $y$, in the Laplace domain, which we denote from now on as~$c_{x,y}^{\mathrm{(Comb)}}(u)$.
To this end, we first isolate $\tilde{c}_{k,y=0}(u)$ in equation~\eqref{eq:correlation fourier}, and take the inverse Fourier transform along the vertical axis $y$ as
\begin{equation}
\tilde{c}_{k,y}(u)=\frac{1}{2\pi}\int_{-\pi}^{\pi}e^{-iqy}\tilde{c}_{k,q}(u)\dd{q}  ,
\end{equation}
finding
\begin{align}
    \tilde{c}_{k,y=0}(u)&=\int_{-\pi}^{\pi}\frac{\dd{q} }{2\pi}\left\lbrace\frac{2(\cos{k}-1)\tilde{c}_{k,y=0}(u)}{u+2(1-\cos{q})}+\frac{\rb(1-\rb)}{u[u+2(1-\cos{q})]}(e^{ik}-1)\right\rbrace \n\\
    &=\frac{2(\cos{k}-1)}{\sqrt{4u+u^{2}}}\tilde{c}_{k,y=0}(u)+\frac{\rb(1-\rb)(e^{ik}-1)}{u\sqrt{4u+u^{2}}}.
    \label{eq:c_along_x}
\end{align}
By taking the inverse Fourier transform along $k$ we then obtain the Laplace transform of the correlations along the backbone
\begin{equation}
    c_{x,y=0}^{\mathrm{(Comb)}}(u)=\frac{\rb(1-\rb)}{u}\int_{-\pi}^{\pi}\frac{\dd{k}}{2\pi}\frac{(e^{ik}-1)e^{-ikx}}{\sqrt{4u+u^{2}}+2(1-\cos{k}) }.
    \label{eq:complex_integral}
\end{equation}
By symmetry of the problem, it is sufficient to calculate $c_{x,y=0}^{\mathrm{(Comb)}}(u)$  for $x\leq 0$, and use
\begin{equation}
    c_{x,y=0}^{\mathrm{(Comb)}}(u)=-c_{-x+1,y=0}^{\mathrm{(Comb)}}(u)
    \label{eq:c_symmetry_x}
\end{equation}
to get the other branch $x> 0$. In fact, the integral in \cref{eq:complex_integral} can be calculated explicitly, yielding for $x\leq 0$
\begin{equation}
   c_{x,y=0}^{\mathrm{(Comb)}}(u)=\frac{\rb(1-\rb)}{2u}
   \frac{\left(\sqrt{u(4+u)}-\sqrt{u^{2}+4[u+\sqrt{u(4+u)}]}\right)}{\sqrt{u^{2}+4[u+\sqrt{u(4+u)}]}}
   \left(1+\frac{\sqrt{u(4+u)}}{2}-\frac{\sqrt{u^{2}+4[u+\sqrt{u(4+u)}]}}{2}\right)^{-x}.
   \label{correlation_laplace_backbone}
\end{equation}

For $y\neq0$ one can either use the expression of $\tilde{c}_{k,y=0}(u)$, implicitly given in~\cref{eq:c_along_x}, into the evolution equation~\eqref{eq:correlation fourier} of the correlations in the Fourier-Laplace domain, or more simply solve the recurrence relation in real space
\begin{equation}
    \begin{cases}
        0=c_{x,y+1}^{\mathrm{(Comb)}}(u) -(u+2)c_{x,y}^{\mathrm{(Comb)}}(u)+
   c_{x,y-1}^{\mathrm{(Comb)}}(u) , & \text{for } y\neq 0,\\
       c_{x,0}^{\mathrm{(Comb)}}(u), & \text{given by \cref{correlation_laplace_backbone},}
    \label{system y>0}
    \end{cases}
\end{equation}
which is obtained by Laplace transforming the bulk equation~\eqref{diff_teeth} for $y\neq0$. Again, it is enough to calculate $c_{x,y}^{\mathrm{(Comb)}}(u)$ for $y>0$ and use the symmetry $ c_{x,y}^{\mathrm{(Comb)}}(u)= c_{x,-y}^{\mathrm{(Comb)}}(u)$ to obtain the correlations in the region $y<0$.
Solving the linear recurrence equation and keeping the solution that does not diverge as $y\to +\infty$ we obtain
\begin{equation}
    \label{eq:c_Comb_Laplace}
   c_{x,y}^{\mathrm{(Comb)}}(u)  =  c_{x,0}^{\mathrm{(Comb)}}(u)\left[1+\frac{u}{2}-\frac{\sqrt{u(4+u)}}{2}\right]^{\abs{y}}.
\end{equation}
    
Finally, the fluctuations of the current can be calculated by inserting the expressions of $ c_{0,0}^{\mathrm{(Comb)}}(u)$ and $c_{1,0}^{\mathrm{(Comb)}}(u)$, given by \cref{eq:c_symmetry_x,correlation_laplace_backbone},
into the evolution equation~\eqref{eq:kappa} of the second cumulant $\kappa(u)$ in the Laplace domain, leading to
\begin{equation}
    \frac{u\kappa(u)}{2}=\frac{\rb(1-\rb)}{u}\sqrt{\frac{u(4+u)}{u^{2}+4[u+\sqrt{u(4+u)} ] }}\n
    \:.
\end{equation}
This expression gives the full time dependence of the fluctuations, up to a Laplace inversion. In particular, the long-time behaviour is obtained from
\begin{equation}
    \frac{u\kappa(u)}{2}\underset{u \to 0^{+}}{\simeq} \frac{\rb(1-\rb)}{\sqrt{2}\, u^{3/4}},
\end{equation}
which using Tauberian theorems~\cite{hughes1995random} gives
\begin{equation}
    \label{eq:FlucCurrentComb}
    \moy{Q_t^2}_{\mathrm{(Comb)}}
    \underset{t \to \infty}{\simeq}
    \rb(1-\rb)\frac{\sqrt{2}}{\Gamma(7/4)}
    t^{3/4}
    \:.
\end{equation}
\begin{figure}
    \centering
    \raisebox{0.05\textwidth}{\includegraphics[height=0.24\textwidth]{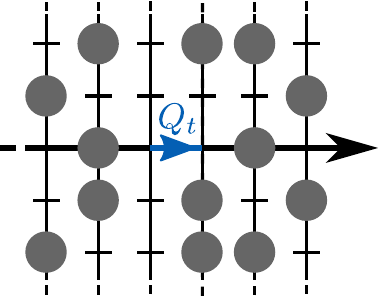}}
    \includegraphics[height=0.3\textwidth]{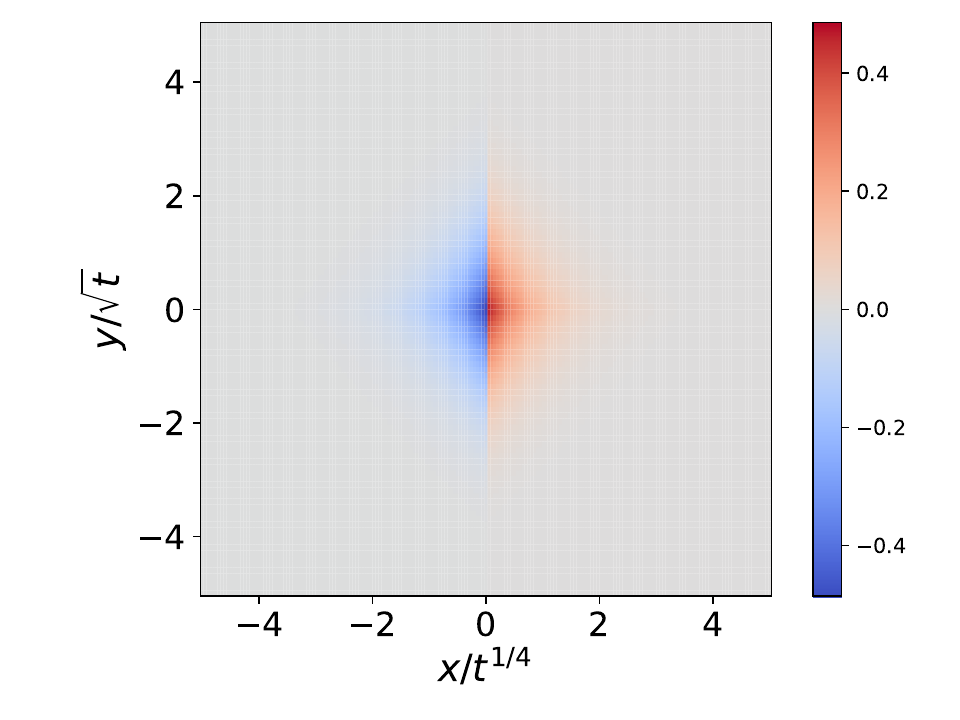}
    \caption{Left: SEP on the comb lattice. On each site, a particle can jump to any neighbouring site ($2$ neighbours on a tooth, $4$ on the backbone) with unit rate. The integrated current $Q_t$ is measured on a link between two sites on the backbone. Right: scaling form of the correlation profile $c_{\vec{r}=(x,y)}^{\mathrm{(Comb)}}(t)/[\bar\rho(1-\bar\rho)]$ for large $t$, obtained from~\eqref{eq:correlation_scaling_comb_Time}.}
    \label{fig:CombQ}
\end{figure}
Remarkably, the current fluctuations behave as $t^{3/4}$ on the comb. This was already observed for reflecting Brownian particles in \cite{Grabsch:2023a}, which corresponds to the low-density limit of \eqref{eq:FlucCurrentComb}. The prefactor of~\eqref{eq:FlucCurrentComb} also matches the one found in the low-density limit~\cite{Grabsch:2023a}. Here, we obtained the full dependence of the prefactor on the density $\bar\rho$, which displays the same functional form $\rb(1-\rb)$ as in the 1D case~\eqref{eq:FlucCurrent1D}. The $t^{3/4}$ scaling shows that the fluctuations are larger on the comb than in the 1D geometry (where they scale as $\sim t^{1/2}$, see \cref{eq:FlucCurrent1D}), but do not grow linearly with time.

This anomalous behavior can be traced back to the correlations $c_{x,y}^{\mathrm{(Comb)}}$ which we have determined at all times in the Laplace domain, see Eq.~\eqref{eq:c_Comb_Laplace}. Its long-time behaviour is encoded in the limit $u\to 0$ at fixed $xu^{1/4}$ and $yu^{1/2}$,
\begin{equation}
    c_{x,y}^{\mathrm{(Comb)}}(u) \underset{u \to 0^{+}}{\simeq}  \sg{x}\frac{ \rb(1-\rb)}{2u}e^{-\sqrt{2}\abs{x}u^{1/4}-\abs{y}u^{1/2}}.
    \label{eq:correlation_scaling_comb_Laplace}
\end{equation}
Taking the inverse Laplace transform yields
\begin{equation}
    c_{x,y}^{\mathrm{(Comb)}}(t) 
    \underset{t \to \infty}{\simeq}  
    \rb(1-\rb) \:
    \mathcal{C}^{\mathrm{(Comb)}} \left( \frac{x}{t^{1/4}}, \frac{y}{\sqrt{t}} \right)
    \:,
    \label{eq:correlation_scaling_comb_Time}
\end{equation}
where we introduced the scaling function
\begin{equation}
    \label{eq:ScalFctComb}
    \mathcal{C}^{\mathrm{(Comb)}}(x,y) = 
    \mathcal{L}^{-1} \left[
    \sg{x}
    \frac{e^{-\sqrt{2}\abs{x}u^{1/4}-\abs{y}u^{1/2}}}{2u} \right](t=1)
    \:,
\end{equation}
defined by the inverse Laplace transform of~\eqref{eq:correlation_scaling_comb_Laplace}, evaluated at $t=1$ (due to the absorption of the $t$ dependence in the scaling variables).
This scaling function is shown in the right panel of Fig.~\ref{fig:CombQ}. Similarly to the 1D case (see Eq.~\eqref{eq:Correl1D}), $c_{x,y}^{\mathrm{(Comb)}}$ generically displays a dipolar structure, i.e.~it assumes positive values in front of the $(0,0)$--$(1,0)$ bond and negative values behind. This is again consistent with the interpretation that higher (lower) currents are correlated with an accumulation (depletion) of particles.
Moreover, correlations spread along $y$ on distances that scale as $\sqrt{t}$, as they do in 1D, meaning that the dynamics within the teeth is essentially the same as in the one-dimensional case. Conversely, correlations spread along the backbone as $t^{1/4}$, which is slower than in 1D; heuristically, this is because particles can get trapped in the teeth, and thus take more time to move along $x$. In this sense, the comb lattice represents an intermediate situation between 1D and 2D lattices.

Finally, these results show that lifting the noncrossing constraint that was imposed in 1$D$ on the dynamics of the particles yields an increase of the \rev{current} fluctuations, but is not sufficient for the system to reach the linear behaviour observed in finite systems~\cite{Bodineau:2008,Akkermans:2013}. 

\subsection{An alternative macroscopic derivation}
Before leaving treelike lattices and moving to 2D and higher dimensions, 
we first show how the current fluctuations~\eqref{eq:FlucCurrentComb} and the correlation profile~\eqref{eq:correlation_scaling_comb_Time} can be alternatively derived from a macroscopic description, which could be applicable to study other models than the SEP, and higher-order cumulants of $Q_t$.

\subsubsection{Macroscopic fluctuation theory on the comb}

The macroscopic fluctuation theory (MFT) provides a general framework to analyse the large scales (long times and large distances) of diffusive systems~\cite{Bertini:2015}, which has permitted to obtain explicit results for various models~\cite{Derrida:2009a,Mallick:2022,Bettelheim:2022,Bettelheim:2022a,Krajenbrink:2022,Grabsch:2024,Grabsch:2024a}.
In the specific case of the SEP, MFT has given exact results both in 1D~\cite{Bodineau:2004,Bodineau:2005,Derrida:2009a,Krapivsky:2014,Krapivsky:2015,Krapivsky:2015a,Grabsch:2022,Grabsch:2023,Mallick:2022,Mallick:2024} and in 2D~\cite{Bodineau:2008,Akkermans:2013,Krapivsky:2014a}.
Since we have seen that the current-density correlation functions are not stationary on the comb, but instead display a scaling behaviour~\eqref{eq:correlation_scaling_comb_Time} indicating that at long times the correlations vary slowly on the scale of the lattice spacing, it is natural to seek a macroscopic description. To the best of our knowledge, the MFT has never been applied to the comb lattice. Therefore we need to adapt the formalism of~\cite{Bertini:2015} to this geometry.

Our starting point is the microscopic equation for the time evolution of the mean occupation of a given site $\vec{r}$, deduced from the microscopic bulk equation~\eqref{eq:profiles_bulk} evaluated at $\lambda = 0$, and which reads
\begin{equation}
    \label{eq:EvolMeanProfComb}
    \partial_t \moy{\eta_{\vec{r}}(t)}
    = 
    \delta_{y,0} \Delta_x \moy{\eta_{\vec{r}}(t)} + \Delta_y \moy{\eta_{\vec{r}}(t)} 
    \:,
    \quad
    \text{for}
    \quad
    \vec{r} = (x,y)
    \:,
\end{equation}
where we have denoted $\Delta_\nu$ the discrete Laplace operator in the direction $\nu = x,y$, i.e. $\Delta_{\nu}f(\vec{r}) = f(\vec{r}+\vec{e}_\nu) - 2 f(\vec{r}) + f(\vec{r}-\vec{e}_\nu)$, with $\vec{e}_\nu$ the unit vector in direction $\nu$. The first term with the Kroenecker delta function $\delta_{y,0}$ enforces that particles can jump horizontally only on the backbone. To derive the macroscopic version of~\eqref{eq:EvolMeanProfComb}, which should hold at long times and large distances, we introduce a large parameter $T$ to rescale the time $t$. Due to the scaling form of the correlations~\eqref{eq:correlation_scaling_comb_Time}, we expect that $x$ scales as $T^{1/4}$ and $y$ as $\sqrt{T}$. Hence, we introduce the macroscopic density $\rho$ as
\begin{equation}
    \label{eq:RescalingDensComb}
    \moy{\eta_{\vec{r}}(t)} 
    \simeq
    \rho \left( \frac{x}{T^{1/4}}, \frac{y}{\sqrt{T}}, \frac{t}{T} \right)
    \:.
\end{equation}
Inserting this scaling form into the evolution equation~\eqref{eq:EvolMeanProfComb}, we get
\begin{equation}
    \partial_t \rho(x,y,t) = \sqrt{T} \delta_{y \sqrt{T},0} \partial_x^2 \rho(x,y,t)
    + \partial_y^2 \rho(x,y,t)
    \:.
\end{equation}
Finally, since $\sqrt{T} \delta_{y \sqrt{T},0} \underset{T \to \infty}{\longrightarrow} \delta(y)$, with $\delta(y)$ the Dirac delta function, we obtain the macroscopic equation for the average density of particles on the comb,
\begin{equation}
    \label{eq:MacroEqComb}
    \partial_t \rho = \delta(y) \partial_x^2 \rho
    + \partial_y^2 \rho
    \:,
\end{equation}
in agreement with e.g.~Eq.~(6) in~\cite{Arkhincheev:2002}.
We can rewrite this equation as a conservation equation with a current $\vec{j}(x,y,t)$, as
\begin{equation}
    \label{eq:ConsEqComb}
    \partial_t \rho + \vec{\nabla} \cdot \vec{j} = 0
    \:,
    \quad
    \text{with}
    \quad
    \vec{j} = - \mathbf{D} \vec{\nabla} \rho
    \:,
    \quad
    \mathbf{D} = \begin{pmatrix}
    \delta(y) & 0\\
    0 & 1
    \end{pmatrix}
    \:.
\end{equation}

By construction, the macroscopic equations~\eqref{eq:ConsEqComb} are deterministic, so they cannot be used to study the fluctuations of density, and subsequently the fluctuations of the integrated current $Q_t$. The idea of the MFT is to postulate a stochastic form for the current in~\eqref{eq:ConsEqComb}, as
\begin{equation}
    \label{eq:StochCurrComb}
    \vec{j} = - \mathbf{D} \vec{\nabla} \rho + 
    \rev{\vec{\zeta}}
    \:,
\end{equation}
with $\vec{\rev{\zeta}}$ a Gaussian white noise, with correlations
\begin{equation}
    \label{eq:CorrEtaComb}
    \moy{\rev{\zeta}_i(x,y,t) \rev{\zeta}_j(x',y',t')} =
    \mathbf{\Sigma}_{i,j}(\rho(x,y,t))
    \delta(x-x')\delta(y-y') \delta(t-t')
    \:,
\end{equation}
and with a correlation matrix $\mathbf{\Sigma}$ depending on the (now stochastic) density $\rho(x,y,t)$. For the 1D SEP, it has been shown that the form~\eqref{eq:StochCurrComb} indeed holds~\cite{Kipnis:1989}, with $\mathbf{\Sigma}(\rho) = 2 \rho(1-\rho)$. In two dimensions, the same relation is expected to hold~\cite{Bertini:2015}, with $\Sigma(\rho) = 2 \rho(1-\rho) \mathds{1}_2$, with $\mathds{1}_2$ the identity matrix. On the comb, we also expect the stochastic form~\eqref{eq:StochCurrComb} of the current to hold, but with the additional constraint that the noise cannot create current fluctuations that make particles jump between teeth, except on the backbone. Therefore, we guess that
\begin{equation}
    \label{eq:SigmaComb}
    \mathbf{\Sigma}(\rho) =
    2 \rho(1-\rho)
    \begin{pmatrix}
    \delta(y) & 0\\
    0 & 1
    \end{pmatrix}
    \:.
\end{equation}
We show below that this assumption gives the correct result for the full distribution of $Q_t$ on the comb in the low-density limit, by retrieving the recent result of~\cite{Grabsch:2023a}. But before that, let us write the MFT equations to study the integrated current $Q_t$.

For this, we follow the general formalism of MFT~\cite{Bertini:2015}, which we briefly summarise here. The first step is to write an action for the fields $\rho$ and $\vec{j}$, using that $\vec{\rev{\zeta}}$ in~\eqref{eq:StochCurrComb} is a Gaussian white noise with correlations~\eqref{eq:CorrEtaComb}, so that its distribution can be formally written as
\begin{equation}
    \label{eq:distrNoiseComb}
    \mathbb{P}[\vec{\rev{\zeta}}] = 
    \exp \left[
        - \frac{1}{2} \int_0^\infty \dd t \int_{-\infty}^\infty \dd x \int_{-\infty}^\infty \dd y
        \: \vec{\rev{\zeta}} \cdot \left( \mathbf{\Sigma}^{-1}(\rho) \vec{\rev{\zeta}} \right)
    \right]
    \:.
\end{equation}
Note that, for the matrix~\eqref{eq:SigmaComb}, the inverse of $\mathbf{\Sigma}$ is singular due to the presence of the delta function. However, this is not a problem since in the final equations we will derive below, only $\mathbf{\Sigma}$ appears and not its inverse. Inserting the relation between the noise and the current~\eqref{eq:StochCurrComb}, and the conservation relation~\eqref{eq:ConsEqComb}, into the distribution~\eqref{eq:distrNoiseComb}, we obtain the joint distribution of $\rho$ and $\vec{j}$,
\begin{equation}
    P[\rho,\vec{j}] =  \exp \left[
        - \frac{1}{2} \int_0^\infty \dd t \int_{-\infty}^\infty \dd x \int_{-\infty}^\infty \dd y
        \: (\vec{j} + \mathbf{D} \vec{\nabla} \rho) \cdot \left( \mathbf{\Sigma}^{-1}(\rho) (\vec{j} + \mathbf{D} \vec{\nabla} \rho)  \right)
    \right]
    \delta(\partial_t \rho + \vec\nabla \cdot \vec{j})
    \:.
\end{equation}
Writing the delta function as an integral over an auxiliary field $H$, and performing the Gaussian integral over the current $\vec{j}$, we obtain
\begin{equation}
    P[\rho] = \int \mathcal{D}H \: \e^{-S[\rho,H]}
    \:,
\end{equation}
with the MFT action~\cite{Bertini:2015},
\begin{equation}
    \label{eq:MFTaction}
    S[\rho,H] = 
    \int_0^\infty \dd t \int_{-\infty}^{+\infty} \dd{x}\rev{\int_{-\infty}^\infty } \dd{y}
    \left[
        H \partial_t \rho + (\mathbf{D} \vec{\nabla}\rho) \cdot \vec{\nabla} H
        - \frac{1}{2} (\mathbf{\Sigma}(\rho) \vec\nabla H) \cdot \vec\nabla H
    \right]
    \:.
\end{equation}
We have left aside up to now for simplicity a key element of MFT, which is the dependence of the amplitude of the noise in~\eqref{eq:CorrEtaComb} on the scaling parameter $T$. When coarse-graining on a large scale, the fluctuations caused by the stochastic microscopic dynamics self-average, so that their amplitude decreases with $T$. In $d$-dimensional systems, the variance of the noise decreases as $T^{-d/2}$~\cite{Bertini:2015}. To identify the scaling on the comb, it is instructive to look at the action~\eqref{eq:MFTaction} and rescale the variables as in~\eqref{eq:RescalingDensComb}, i.e.~$x \to x/T^{1/4}$, $y \to y/\sqrt{T}$, $t \to t/T$. This gives
\begin{equation}
    S[\rho,H] \to T^{3/4} \: S[\rho,H]
    \:,
\end{equation}
indicating that the variance of the macroscopic noise $\vec{\rev{\zeta}}$ decreases as $T^{-3/4}$ on the comb. Finally, the correct form of the distribution of the macroscopic density reads
\begin{equation}
    \label{eq:MacroDistRhoComb}
    P[\rho] = \int \mathcal{D}H \: \e^{-T^{3/4} \: S[\rho,H]}
    \:,
    \quad \text{with} \quad
    \rho(x,y,t) \simeq \eta_{(x T^{1/4}, y \sqrt{T})}(t T)
    \:.
\end{equation}
This provides a macroscopic description of the stochastic density of particles for the SEP on the comb, in the framework of MFT. The last ingredient needed is the initial distribution of particles. As discussed above, we start from an equilibrium distribution, which for the comb corresponds to populating each site independently, with probability $\rb$, as shown in Eq.~\eqref{equilibrium_distribution}. This is identical to the equilibrium measure for the SEP in 2D, which at the macroscopic level becomes~\cite{Derrida:2007}
\begin{equation}
    P[\rho(x,y,0)] = \e^{- T^{3/4} \: F[\rho(x,y,0)]}
    \:,
    \quad
    F[\rho(x,y,0)] = \int_{-\infty}^\infty \dd{x} \rev{\int_{-\infty}^\infty } \dd{y} 
    \int_{\rb}^{\rho(x,y,0)} \frac{\rho(x,y,0) - z}{z(1-z)} \dd{z}
    \:.
    \label{eq:F_comb_MFT}
\end{equation}
We can now apply this MFT formalism to the study of the integrated current.

\subsubsection{Application to the integrated current}
\label{sec:int_current_comb_mft}

Since the comb lattice has a tree structure, the integrated current through the bond $(0,0)-(1,0)$ up to time $T$ is equal to the variation of the number of particles in the domain $x > 0$, which thus reads
\begin{equation}
    Q_T = \sum_{x > 0} \sum_{y \in \mathbb{Z}} [\eta_{(x,y)}(T) - \eta_{(x,y)}(0) ]
    \simeq T^{3/4} \int_0^\infty \dd x \int_{-\infty}^{\infty} \dd{y}
    [\rho(x,y,1) - \rho(x,y,0)]
    \:,
\end{equation}
where we used the scaling form~\eqref{eq:MacroDistRhoComb}. Having expressed $Q_T$ in terms of the macroscopic density $\rho$, we can follow the same approach as in 1D~\cite{Derrida:2009a} and write the moment generating function of $Q_T$ as
\begin{equation}
    \label{eq:CumulCombMFT}
    \moy{\e^{\lambda Q_T}} \simeq \int \mathcal{D}\rho \mathcal{D}H \:
    \exp \left[- T^{3/4} \left(
    S[\rho,H] + F[\rho(x,y,0)] - \lambda \int_0^\infty \dd x \int_{-\infty}^{\infty} \dd{y}
    [\rho(x,y,1) - \rho(x,y,0)]
    \right)
    \right]
    \:.
\end{equation}
Evaluating this integral in the long-time limit $T \to \infty$ by the saddle-point method, we get
\begin{equation}
    \label{eq:MGFCombMFT}
    \moy{\e^{\lambda Q_T}} \underset{T \to \infty}{\simeq}
    \e^{T^{3/4} \hat{\psi}(\lambda)}
    \:,
    \quad
    \hat\psi(\lambda) = 
    \lambda \int_0^\infty \dd x \int_{-\infty}^{\infty} \dd{y}
    [q(x,y,1) - q(x,y,0)]
    - S[q,p] - F[q(x,y,0)]
    \:,
\end{equation}
where we have denoted $(q,p)$ the saddle point for $(\rho,H)$, which obeys the saddle-point equations:
\begin{align}
\label{eq:MFTqComb}
    \partial_t q 
    &= \delta(y) \partial_x^2 q + \partial_y^2 q
    - \delta(y) \partial_x[2q(1-q) \partial_x p]
    - \partial_y[2q(1-q) \partial_y p]
    \:,
    \\
\label{eq:MFTpComb}
    \partial_t p
    &=
    - \delta(y) \partial_x^2 p - \partial_y^2 p
    - (1-2q) ( \delta(y) (\partial_x p)^2 + (\partial_y p)^2 )
    \:,
\end{align}
with the boundary conditions
\begin{equation}
\label{eq:BondCondMFTComb}
    p(x,y,0) = \lambda \Theta(x) + \int_{\rb}^{q(x,y,0)} \frac{\dd z}{z(1-z)}
    \:,
    \quad
    p(x,y,1) = \lambda \Theta(x)
    \: ,
\end{equation}
\rev{and where $\Theta(x)$ is the Heaviside theta function.}
Note that~\eqref{eq:MGFCombMFT} implies that the cumulant generating function $\ln \moy{\e^{\lambda Q_T}}$ scales as $T^{3/4}$. In particular, at order $2$ in $\lambda$, we recover the correct scaling of the variance~\eqref{eq:FlucCurrentComb}.

From the solution of the MFT equations~\eqref{eq:MFTqComb} to~\eqref{eq:BondCondMFTComb}, the cumulants of $Q_T$ are then obtained by expanding the cumulant generating function $\hat\psi$~\eqref{eq:MGFCombMFT} in powers of $\lambda$. In practice, it is actually simpler to compute them through the derivative
\begin{equation}
    \label{eq:DpsiDlambdaComb}
    \frac{\dd \hat{\psi}}{\dd \lambda} =
     \int_0^\infty \dd x \int_{-\infty}^{\infty} \dd y
    [q(x,y,1) - q(x,y,0)]
    \:,
\end{equation}
since $(q,p)$ is the saddle point of $S+F-\lambda Q_T$.

Furthermore, this MFT approach also gives access to the correlations between the integrated current $Q_T$ and the density of particles $\rho(x,y,t)$, which are encoded in the correlation profiles~\eqref{correlation_comb}, which can be computed in MFT as~\cite{Poncet:2021}
\begin{equation}
    \label{eq:CorrProfMFT}
    \frac{\moy{\eta_{\vec{r}=(x,y)}(T) \e^{\lambda Q_T}}}{\moy{\e^{\lambda Q_T}}}
    \simeq
    \frac{
    \int \mathcal{D}\rho \mathcal{D}H \:
    \rho(x,y,1) \: \e^{- T^{3/4} \left(
    S[\rho,H] + F[\rho(x,y,0)] - \lambda \int_0^\infty \dd x \int_{-\infty}^{\infty} \dd y
    [\rho(x,y,1) - \rho(x,y,0)]
    \right)}
    }{
    \int \mathcal{D}\rho \mathcal{D}H \:
    \e^{- T^{3/4} \left(
    S[\rho,H] + F[\rho(x,y,0)] - \lambda \int_0^\infty \dd x \int_{-\infty}^{\infty} \dd y
    [\rho(x,y,1) - \rho(x,y,0)]
    \right)
    }
    }
    \underset{T \to \infty}{\simeq}
    q(x,y,1)
    \:.
\end{equation}
The correlation profiles thus coincide with the solution of the MFT equations~(\ref{eq:MFTqComb}-\ref{eq:BondCondMFTComb}) at final time $t=1$. In turn, expanding this solution in powers of $\lambda$ gives the long-time behaviour of the correlations, for instance
\begin{equation}
    q(x,y,1)  \underset{T \to \infty}{\simeq}
    \moy{\eta_{\vec{r}}(T)}
    + \lambda \Big( 
    \moy{\eta_{\vec{r}}(T) Q_T} - \moy{\eta_{\vec{r}}(T)} \moy{Q_T} 
    \Big)
    + \mathcal{O}(\lambda^2)
    \:.
\end{equation}
Thus, we only need to compute the first order in $\lambda$ of $q$ to deduce the correlation profile $c_{\vec{r}}^{\mathrm{(Comb)}}(T)$ and the current fluctuations $\moy{Q_T^2}_{\mathrm{(Comb)}}$. But first we show that the MFT equations, derived from the guess on the stochastic current~(\ref{eq:StochCurrComb},\ref{eq:SigmaComb}), reproduce known results in the low-density limit~\cite{Grabsch:2023a}, which thus validates them.

\subsubsection{Solution for the full distribution in the low-density limit}

In 1D, the SEP at low density reduces to a system of independent Brownian particles~\cite{Imamura:2017}. This system has recently been studied on the comb~\cite{Grabsch:2023a}, and provides a benchmark for the MFT equations~(\ref{eq:MFTqComb}-\ref{eq:BondCondMFTComb}). Keeping only the leading order in $q$, Eqs.~(\ref{eq:MFTqComb}-\ref{eq:BondCondMFTComb}) reduce to
\begin{align}
\label{eq:MFTqCombLowDens}
    \partial_t q 
    &= \delta(y) \partial_x^2 q + \partial_y^2 q
    - \delta(y) \partial_x[2q \partial_x p]
    - \partial_y[2q \partial_y p]
    \:,
    \\
\label{eq:MFTpCombLowDens}
    \partial_t p
    &=
    - \delta(y) \partial_x^2 p - \partial_y^2 p
    - \delta(y) (\partial_x p)^2 - (\partial_y p)^2
    \:,
\end{align}
with the boundary conditions
\begin{equation}
\label{eq:BondCondMFTcombLowDens}
    p(x,y,0) = \lambda \Theta(x) 
    +
    \ln \left( \frac{q(x,y,0)}{\rb} \right)
    \:,
    \quad
    p(x,y,1) = \lambda \Theta(x)
    \:.
\end{equation}
These equations can be solved, for arbitrary $\lambda$, by the same Cole-Hopf transform as in the 1D case~\cite{Derrida:2009a}. We first introduce
\begin{equation}
    \label{eq:ColeHopfComb}
    Q = q \e^{-p}
    \:,
    \quad
    P = \e^p
    \:.
\end{equation}
Inserting this change of functions into~(\ref{eq:MFTqCombLowDens}-\ref{eq:BondCondMFTcombLowDens}), we obtain that $Q$ and $P$ satisfy simple diffusion and anti-diffusion equations on the comb,
\begin{equation}
    \partial_t Q =  \delta(y) \partial_x^2 Q + \partial_y^2 Q
    \:,
    \quad
    \partial_t P = -\delta(y) \partial_x^2 P - \partial_y^2 P
    \:,
\end{equation}
with the boundary conditions
\begin{equation}
    Q(x,y,0) = \rb \: \e^{- \lambda \Theta(x)}
    \:,
    \quad
    P(x,y,1) = \e^{\lambda \Theta(x)}
    \:.
\end{equation}
The solution of these equations can be expressed in terms of the Green's function of the diffusion equation on the comb, $G_t(x,y|x',y')$, given explicitly in the Laplace domain by (see Appendix~\ref{app:GreenFctComb})
\begin{equation}
    \label{eq:PropComb}
    \int_0^\infty \e^{-s t} G_t(x,y|x',y') \dd t
    = \frac{1}{2 \sqrt{2} \: s^{1/4}}
    \e^{- \sqrt{2} s^{1/4} \abs{x-x'} - \sqrt{s}(\abs{y} + \abs{y'})}
    + \frac{1}{2 \sqrt{s}} \delta(x-x') \Theta(yy')
    \left(
    \e^{-\sqrt{s} \abs{y-y'}}
    - \e^{-\sqrt{s} \abs{y+y'}}
    \right)
    \:.
\end{equation}
Such solution reads,
\begin{equation}
    Q(x,y,t) = \rb  \int_{-\infty}^\infty \dd x' \dd y' G_t(x,y|x',y') \e^{-\lambda \Theta(x')}
    \:,
    \quad
    P(x,y,t) = \int_{-\infty}^\infty \dd x' \dd y' G_{1-t}(x,y|x',y') \e^{\lambda \Theta(x')}
    \:.
\end{equation}
Going back using~\eqref{eq:ColeHopfComb} to the solution $q$ giving the correlation profiles~\eqref{eq:CorrProfMFT}, we finally get
\begin{equation}
    q(x,y,1) = \rb \: \e^{\lambda \Theta(x)} \int_{-\infty}^\infty \dd x' \dd y' G_1(x,y|x',y') \e^{-\lambda \Theta(x')}
    \:,
\end{equation}
which coincides with the result of Ref.~\cite{Grabsch:2023a}, thus validating our MFT approach on the comb.

\subsubsection{Solution for the fluctuations at arbitrary density}

We now turn to the determination of the fluctuations of $Q_T$ and the associated correlations for the SEP, at arbitrary density. Expanding the saddle-point solution $(q,p)$ in powers of $\lambda$,
\begin{equation}
    q = \rb + \lambda q_1 + \mathcal{O}(\lambda^2)
    \:,
    \qquad
    p = \lambda p_1 + \mathcal{O}(\lambda^2)
    \:,
\end{equation}
we obtain the MFT equations at first order,
\begin{align}
\label{eq:MFTqCombOrder1}
    \partial_t q_1
    &= \delta(y) \partial_x^2 q_1 + \partial_y^2 q_1
    - 2 \rb(1-\rb) \left[ \delta(y) \partial_x^2 p
    + \partial_y^2 p \right]
    \:,
    \\
\label{eq:MFTpCombOrder1}
    \partial_t p_1
    &=
    - \delta(y) \partial_x^2 p_1 - \partial_y^2 p_1
    \:,
\end{align}
with the boundary conditions
\begin{equation}
\label{eq:BondCondMFTCombOrder1}
    p_1(x,y,0) = \Theta(x) + \frac{q_1(x,y,1)}{\rb(1-\rb)}
    \:,
    \quad
    p_1(x,y,1) = \Theta(x)
    \:.
\end{equation}
The equation for $p_1$ can be solved explicitly using the Green's function~\eqref{eq:PropComb},
\begin{equation}
    \label{eq:p1MFTcomb}
    p_1(x,y,t) = \int_0^\infty \dd x' \int_{-\infty}^\infty \dd y'
    G_{1-t}(x,y|x',y')
    \:.
\end{equation}
Using the expression of the propagator in the Laplace domain~\eqref{eq:PropComb}, 
$p_1$ can be written as an inverse Laplace transform,
\begin{equation}
\label{eq:p1InvLap}
    p_1(x,y,t)
    = \Theta(x) - \mathcal{L}^{-1} \left[\frac{1}{2s}
    \sg{x} \e^{-\sqrt{2} s^{1/4} \abs{x} - \sqrt{s} \abs{y}}
    \right](1-t)
    \:.
\end{equation}
The equation for $q_1$ can be solved similarly by noticing that the change of functions
\begin{equation}
    q_1 = \rb(1-\rb) p_1 + \tilde{q}_1
\end{equation}
yields an equation for $\tilde{q}_1$,
\begin{equation}
    \partial_t \tilde{q}_1 = \delta(y) \partial_x^2 \tilde{q}_1 + \partial_y^2 \tilde{q}_1
    \:,
    \quad
    \tilde{q}_1(x,y,0) = - \rb(1-\rb) \Theta(x)
    \:,
\end{equation}
which is identical to the equation for $p_1$, up to rescaling by $-\rb(1-\rb)$ and changing $t \to 1-t$. Therefore, the solution for $q_1$ reads
\begin{equation}
    \label{eq:Solq1Comb}
    q_1(x,y,t) = \rb(1-\rb) \left[
    p_1(x,y,t) - p_1(x,y,1-t)
    \right]
    \:.
\end{equation}
In particular, at final time $t=1$, this gives the correlation profile
\begin{equation}
    c^{\mathrm{(Comb)}}_{\vec{r}=(x T^{1/4},y \sqrt{T})}(T) = \mathrm{Cov}(Q_T, \eta_{\vec{r}}(T))
    \underset{T \to \infty}{\simeq}
    q_1(x,y,1) = \rb(1-\rb) \left[
    \Theta(x) - p_1(x,y,0)
    \right]
    \:.
\end{equation}
Using the expression of $p_1$~\eqref{eq:p1InvLap} yields
\begin{equation}
    c^{\mathrm{(Comb)}}_{\vec{r}=(x T^{1/4},y \sqrt{T})}(T) \underset{T \to \infty}{\simeq}
    \rb(1-\rb) 
    \mathcal{L}^{-1} \left[\frac{1}{2s}
    \sg{x} \e^{-\sqrt{2} s^{1/4} \abs{x} - \sqrt{s} \abs{y}}
    \right](t=1)
    \:,
\end{equation}
which coincides with the expression~\eqref{eq:correlation_scaling_comb_Time} obtained from microscopic calculations. This further supports the validity of the MFT description of the SEP on the comb.

The fluctuations of $Q_T$ can then by deduced using~\eqref{eq:DpsiDlambdaComb} at order $1$ in $\lambda$, which together with~\eqref{eq:Solq1Comb} yields
\begin{equation}
    \moy{Q_T^2}_{\mathrm{(Comb)}}
    \underset{T \to \infty}{\simeq} T^{3/4}
    \: 2 \rb(1-\rb) \int_0^\infty \dd x \int_{-\infty}^\infty \dd y
    \left[
        1 - p_1(x,y,0)
    \right]
    \:.
\end{equation}
Using the expression of $p_1$~\eqref{eq:p1InvLap}, the fluctuations can be computed explicitly, and read
\begin{equation}
    \moy{Q_T^2}_{\mathrm{(Comb)}}
    \underset{T \to \infty}{\simeq}
    \rb(1-\rb) \frac{\sqrt{2}}{\Gamma(\frac{7}{4})} \:  T^{3/4}
    \:,
\end{equation}
reproducing again the microscopic result~\eqref{eq:FlucCurrentComb}.

Unlike the microscopic formalism which does not yield closed equations for higher-order cumulants of $Q_T$ and the associated correlation profiles, the MFT equations~\eqref{eq:MFTqComb}~to~\eqref{eq:BondCondMFTComb} obtained here are closed, and open the way to obtaining the full distribution of $Q_T$ for the SEP on the comb. This method could also be applied to other models than the SEP on the comb, as it was done for the usual MFT in 1D or higher dimensions~\cite{Bertini:2015}.

\section{2D and higher-dimensional lattices}
\label{sec:2D}

As we have seen by considering the comb lattice, lifting the noncrossing condition is not sufficient for the correlations to reach a steady state at long times, and thus yield a linear behaviour for the fluctuations of $Q_t$. We now lift the second dynamical constraint by considering looped lattices. A single particle can now give an arbitrarily high contribution to $Q_t$ by looping several times through the bond $(0,0)-(1,0)$.

As for the comb, we first derive the correlation profiles using a microscopic approach based on the master equation of the process, from which we deduce the current fluctuations. We then pinpoint the key role played by the looped structure of the lattice by showing that the main contribution to $Q_t$ comes from particles
that circulate repeatedly through the bond $(0,0)-(1,0)$, thus giving rise to
\textit{vortices} in the current profile.

\subsection{Fluctuations and correlation profiles in \texorpdfstring{$D=2$}{D=2}}
\label{sec:MasterEq2D}

\begin{figure}
    \centering
    \raisebox{0.05\textwidth}{\includegraphics[height=0.24\textwidth]{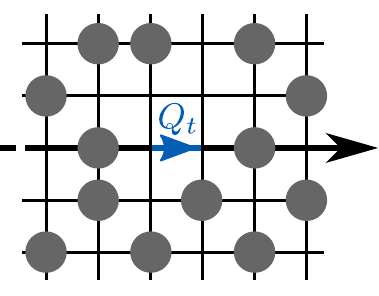}}
    \includegraphics[height=0.3\textwidth]{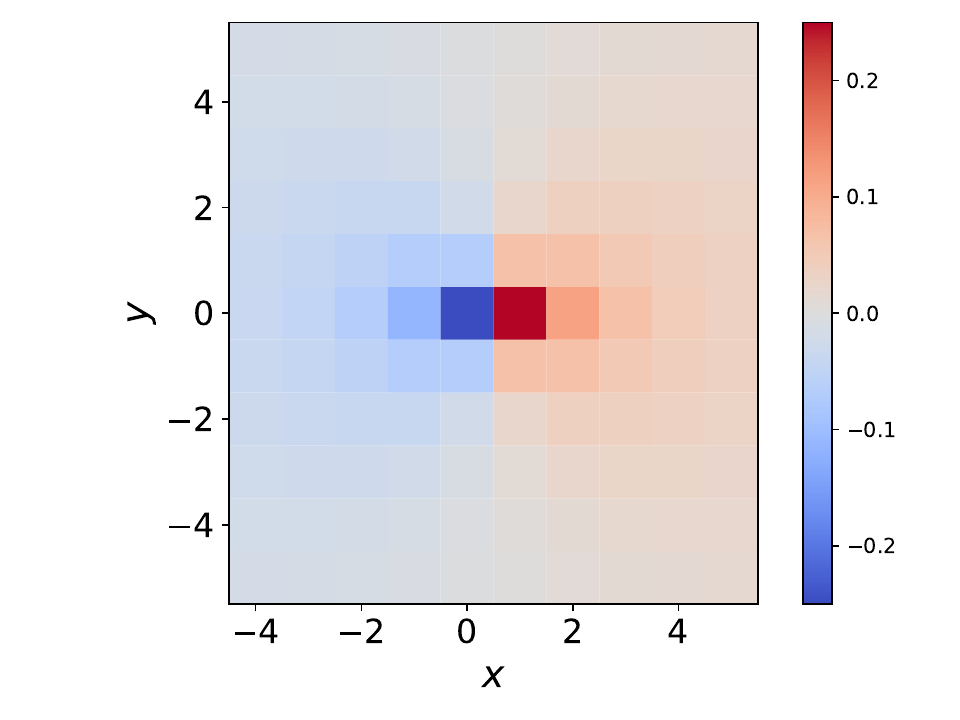}
    \caption{Left: SEP on a 2D lattice. The integrated current $Q_t$ is measured on the bond between the sites $(0,0)$ and $(1,0)$. Right: correlation profile $c_{\vec{r}=(x,y)}^{\mathrm{(2D)}}(t)/[\bar\rho(1-\bar\rho)]$ for large $t$, obtained from~\eqref{eq:CorrProf2D}. }
    \label{fig:2DQ}
\end{figure}

In 2D, the microscopic derivation of the current fluctuations and the correlation profiles follows a similar procedure to  the one we adopted for the comb in~\cref{comb_microscopic}. In the 2D case the master equation is given by
\begin{equation}
\label{eq:MasterEq2D}
    \partial_t P_t(\underline{\eta})
    = \sum_{x, y} \left[P(\underline{\eta}^{x,y+}, t) - P(\underline{\eta}, t) \right] + \sum_{x,y} \left[P(\underline{\eta}^{x+,y}, t) - P(\underline{\eta}, t) \right]  ,
\end{equation}
where we kept the same notations as in \cref{comb_microscopic}. One can see that the uniform distribution in \cref{equilibrium_distribution} also represents the equilibrium state in 2D.
    Using the master equation~\eqref{eq:MasterEq2D} we first obtain the evolution equation verified by $\psi$ introduced in~\cref{generating_function_comb}, namely
\begin{equation}
    \partial_t\psi(\lambda, t) = (e^{\lambda}-1)\frac{\moy{\eta_{0,0}(1-\eta_{1,0})e^{\lambda Q_t}}}{\moy{e^{\lambda Q_t}}} + (e^{-\lambda}-1)\frac{\moy{\eta_{1,0}(1-\eta_{0,0})e^{\lambda Q_t}}}{\moy{e^{\lambda Q_t}}}.
     \label{eq:cumulant-2D}
\end{equation}
Let us now consider the evolution of $w_{x,y}$ introduced in~\cref{correlation_comb}. As for the comb, only an exchange of occupations between sites $(0,0)$ and $(1,0)$ can affect the current, and thus it is convenient to consider separately those sites from the sites in the bulk.
Using the master equation~\eqref{eq:MasterEq2D} we get for the bulk, i.e. for~$(x,y)\neq \{(0,0), (1,0)\}$,
\begin{multline}
    \partial_tw_{x,y} =\sum_{\mu = \pm1}\nabla_{\mu}^yw_{x,y} \quad + \quad \sum_{\mu = \pm1}\nabla_{\mu}^xw_{x,0}   \\ + \quad (e^{\lambda}-1)\left[\frac{\moy{\eta_{ x,y}\eta_{ 0,0}(1-\eta_{1,0})e^{\lambda Q_t}}}{\rev{\moy{e^{\lambda Q_t}}}} - w_{x,y}\frac{\moy{\eta_{0,0}(1-\eta_{1,0})e^{\lambda Q_t}}}{ \rev{\moy{e^{\lambda Q_t}}} } \right] \\
    + \quad (e^{-\lambda}-1)\left[\frac{\moy{\eta_{ x,y}\eta_{ 1,0}(1-\eta_{0,0})e^{\lambda Q_t}}}{\rev{\moy{e^{\lambda Q_t}}}} - w_{x,y}\frac{\moy{\eta_{1,0}(1-\eta_{0,1})e^{\lambda Q_t}}}{\rev{\moy{e^{\lambda Q_t}}}} \right],
    \label{eq:profiles_bulk2D}
\end{multline}
and for the boundaries
\begin{align}
     &\partial_tw_{0,0} =\biggl[e^{-\lambda} - (e^{-\lambda}-1)w_{0,0} \biggr]\frac{\partial_t\psi(\lambda, t)}{e^{-\lambda}-1} +\sum_{\mu=\pm1}\nabla_{\mu}^yw_{0,0} \quad + \nabla_{-1}^xw_{0,0} ,\label{masterboundary02d} \\
     &\partial_tw_{1,0} =\biggl[e^{\lambda} - (e^{\lambda}-1)w_{1,0} \biggr]\frac{\partial_t\psi(\lambda, t)}{e^{\lambda}-1} +\sum_{\mu=\pm1}\nabla_{\mu}^yw_{1,0} \quad + \nabla_{1}^xw_{1,0} ,
     \label{masterboundary12d}
\end{align}
where we have used~\cref{eq:cumulant-2D}.
Expanding the evolution equation of the generating function~\eqref{eq:cumulant-2D} up to order 2 in $\lambda$ we find
\label{sec:FlucJ2D}
\begin{equation}
\frac{\partial_t\moy{Q_t^{2}}}{2}= c_{0,0}^{\mathrm{(2D)}}(t)-c_{1,0}^{\mathrm{(2D)}}(t) +\rb(1-\rb), \label{fluctuation_laplace_2D}
\end{equation}
where, analogously to \cref{eq:c_comb}, we have defined
\begin{equation}
    c_{\vec{r}=(x,y)}^{\mathrm{(2D)}}(t) \equiv \moy{\eta_{\vec{r}}Q_t}_c = \moy{\eta_{\vec{r}}Q_t}
    \label{correlation_2D}
\end{equation}
(indeed, $\moy{Q_t}=0$).
At equilibrium, correlations between sites vanish up to order 1 in $\lambda$, so that at leading order we get from \cref{eq:profiles_bulk2D,masterboundary02d,masterboundary12d}
\begin{align}
  &\partial_t c_{\vec{r}}^{\mathrm{(2D)}}(t) = \sum_{\mu = \pm1}\nabla_{\mu}^y c_{\vec{r}}^{\mathrm{(2D)}}(t)+\sum_{\mu = \pm1}\nabla_{\mu}^x c_{\vec{r}}^{\mathrm{(2D)}}(t),
    \label{diff_2D}\\
     &\partial_t c_{0,0}^{\mathrm{(2D)}}(t) = -\frac{\partial_t\moy{Q_t^{2}}}{2}+ \sum_{\mu = \pm1}\nabla_{\mu}^y c_{0,0}^{\mathrm{(2D)}}(t)+ \nabla_{-1}^x c_{0,0}^{\mathrm{(2D)}}(t),
     \label{boundary2D_0}\\
      &\partial_t c_{1,0}^{\mathrm{(2D)}}(t) = \frac{\partial_t\moy{Q_t^{2}}}{2}+ \sum_{\mu = \pm1}\nabla_{\mu}^y c_{1,0}^{\mathrm{(2D)}}(t)+ \nabla_{1}^x c_{1,0}^{\mathrm{(2D)}}(t).
      \label{boundary2D_1}
\end{align}
Equations~\eqref{fluctuation_laplace_2D}~and~\eqref{diff_2D}~to~\eqref{boundary2D_1} now form a closed set, which can be solved in the Fourier-Laplace domain by introducing  
\begin{equation}
\tilde{c}_{k,q}(u)= \sum_{x,y}\int_{0}^{+\infty}e^{-ut}e^{ikx+iqy} c_{\vec{r}=(x,y)}^{\mathrm{(2D)}}(t)\dd{t}.
\end{equation}
Applying the Laplace transform to~\cref{fluctuation_laplace_2D} and the Fourier-Laplace transform to \cref{diff_2D,boundary2D_0,boundary2D_1} we now obtain
\begin{align}
 &u\frac{\kappa(u)}{2}= \frac{\rb(1-\rb)}{u}+ c_{0,0}(u) - c_{1,0}(u),\\
 &  u\, \tilde{c}_{k,q}(u)=2(\cos{q}-1)\tilde{c}_{k,q}(u)+2(\cos{k}-1)\tilde{c}_{k,q}(u)+\frac{\rb(1-\rb)}{u}(e^{ik}-1).
 \label{fourier_correlation_2D}
\end{align}
In contrast to the case of the comb, the relation~\eqref{fourier_correlation_2D} verified by $\tilde{c}_{k,q}(u)$ does not involve other unknown quantities, and then we can deduce directly the correlations in real space (but still in the Laplace domain) by inverse Fourier transform:
\begin{equation}
   c_{\vec{r}=(x,y)}^{\mathrm{(2D)}}(u)=\frac{\rb(1-\rb)}{u}\iint_{-\pi}^{\pi}\frac{\dd{k}}{2\pi}\frac{\dd{q}}{2\pi}\frac{(e^{ik}-1)e^{-ikx -iqy}}{u+2(1-\cos{q})+2(1-\cos{k})}.
\end{equation}
This expression holds for all $u$, and thus at all times $t$ by inverse Laplace transform. Contrary to the case of the comb, in which  $c_{\vec{r}=(x,y)}^{\mathrm{(Comb)}}(u)$ assumes the scaling form~\eqref{eq:correlation_scaling_comb_Laplace}, here $c_{\vec{r}=(x,y)}^{\mathrm{(2D)}}(u)$ converges to a finite value, indicating that the correlations in the time domain converge to
\begin{equation}
    \label{eq:CorrProf2D}
    c_{\vec{r}=(x,y)}^{\mathrm{(2D)}}(t)
    \underset{t \to \infty}{\simeq} \frac{\rb(1-\rb)}{2}\iint_{-\pi}^{\pi}\frac{\dd{k}}{2\pi}\frac{\dd{q}}{2\pi}\frac{(e^{ik}-1)e^{-ikx -iqy}}{(1-\cos{q})+(1-\cos{k})}.
\end{equation}
This correlation profile is shown in Fig.~\ref{fig:2DQ}. It still displays a dipolar structure, with positive correlations for $x \geq 1$ and negative correlations for $x \leq 0$, indicating that a fluctuation that increases $Q_t$ corresponds to an increase of the density on the right of the bond $(0,0)-(1,0)$, and a decrease on its left. However, unlike the cases of 1D~\eqref{eq:Correl1D} or comb geometry~\eqref{eq:correlation_scaling_comb_Time} (in which the correlations spread with time and thus vary slowly at the scale of the lattice spacing), the correlations~\eqref{eq:CorrProf2D} in 2D are stationary and display important variations between neighbouring lattice sites. Such structure cannot be captured by macroscopic approaches like MFT.

Finally, the fact that the correlations~\eqref{eq:CorrProf2D} are stationary yields that the fluctuations of the integrated current $Q_t$, which can be deduced from \cref{fluctuation_laplace_2D} as
\begin{equation}
    \label{eq:FlucCurrent2D}
    \moy{Q_t^2}_{\mathrm{(2D)}}
    \underset{t \to \infty}{\simeq}
    2\rb(1-\rb)\left[1+\frac12 \iint_{-\pi}^{\pi}\frac{\dd{k}}{2\pi}\frac{\dd{q}}{2\pi}\frac{(e^{ik}-1)(1-e^{-ik})}{2(1-\cos{q})+2(1-\cos{k})}\right]t 
    =  \rb(1-\rb)t
    \:,
\end{equation}
grow linearly with time as in the case of finite systems.

To show that the looped structure of the lattice plays a central role in the fact that the correlations in this infinite system reach a steady state, we now turn to an alternative formalism.
This will give additional insight on the behavior of the currents in higher dimensions.

\subsection{Microscopic calculation in higher dimensions}
\label{sec:micro_action}

To identify the physical origin of the stationary correlation profile~\eqref{eq:CorrProf2D}, we now analyse the current-current correlations. Since the correlation profile~\eqref{eq:CorrProf2D} is stationary, there is no natural large length scale in the long-time limit that can be used to build a continuous description like the MFT to analyse these currents, 
which forces us to
keep a discrete formalism. A similar approach was used in~\cite{Lefevre:2007,Saha:2023} for the 1D SEP.

\subsubsection{Constructing a microscopic action formalism}
\label{sec:micro_high_D}

We start by 
breaking the total measurement time $T$ into $M$ steps of duration $\dd t$.
Here and henceforth, we will call by convention $\nuv$ each one of the basis vectors of the lattice $\{ \vec{e}_1, \ldots, \vec{e}_D \}$, while $\muv \in \{ \pm \vec{e}_1, \ldots, \pm \vec{e}_D \}$.
We then
write the microscopic conservation relation on each site
$\vec{r}$,
\begin{equation}
    \eta_{\vec{r}}(t+\dd t) - \eta_{\vec{r}}(t)
    = \dd t \sum_{\nuv}
    \left[ \vec{j}_{\rv - \nuv}(t) - \vec{j}_{\rv }(t) \right] \cdot \nuv
    \:,
    \label{eq:multi_eta}
\end{equation}
where $\vec{j}_{\rv}(t)$ is the current from site $\rv$ at time $t$, so that $\vec{j}_{\rv}(t) \cdot \nuv \dd t$ is the number of particles that have left the site in direction $\nuv$ during $\dd t$.
The current through each bond can be described as
\begin{equation}
    \vec{j}_{\rv}(t) \dd t
    = \sum_{\nuv} \left[
    \eta_{\rv} (1-\eta_{\rv+\nuv}) \xi_{\vec{r},\nuv}(t)
    - \eta_{\rv+\nuv} (1-\eta_{\rv}) \xi_{\vec{r}+\nuv,-\nuv}(t)
    \right]
    \nuv
    \:,
    \label{eq:multi_j}
\end{equation}
where $\xi_{\rv,\muv}(t)$ is a Poissonian random variable determining whether a particle located at $\rv$ at time $t$ jumps in direction $\muv$ in $[t,t+\dd t]$, namely
\begin{equation}
    \xi_{\rv,\muv}(t) =
    \left\lbrace
    \begin{array}{ll}
        1 & \text{with probability } \gamma \dd t \:,  \\
        0 & \text{with probability } 1-\gamma \dd t \:.
    \end{array}
    \right.
    \label{eq:multi_xi}
\end{equation}
These stochastic equations fully determine the dynamics of the system, and are the microscopic equivalent of~(\ref{eq:ConsEqComb},\ref{eq:StochCurrComb}). 
Note that, unlike what is usually done in the SEP, here the exponential ``clocks'' are not assumed to be in correspondence of the particles, but rather on each 
bond
of the lattice.
The rate $\gamma$ can be related to the jump rate of the particles as follows: if a particle jumps with probability
$\dd t/\tau^\star$ in any of the $2D$ directions, it means that it jumps in a given direction with probability $\dd
t/(2D \tau^\star)$, and therefore the rate $\gamma = 1/(2D \tau^\star)$. (In the previous sections, we have considered $\gamma=1$.) It is simple to check that \cref{eq:multi_eta,eq:multi_j,eq:multi_xi} lead to the standard master equation for the occupations in SEP, i.e.
\begin{equation}
  P_{t+\dd t}[ \underline{\eta}] =
  P_t[\underline{\eta}]
  + \gamma \,\dd t \sum_{\vec{r}} \sum_{\vec{\mu}}
  \left(
  \eta_{\vec{r}+\muv}(1 - \eta_{\vec{r}}) P_t \left[ \underline{\eta}^{\vec{r},\muv}\right]
  - \eta_{\vec{r}}(1 - \eta_{\vec{r}+\muv}) P_t[\underline{\eta}]
  \right),
\end{equation}
where we have denoted as $\underline{\eta}^{\vec{r},\vec{\mu}}$ the
configuration in which the occupations of the sites at positions
$\vec{r}$ and $\vec{r}+\vec e_\mu$ have been exchanged. 

Next, \cref{eq:multi_eta,eq:multi_j,eq:multi_xi} can be used to build an action formalism, at the microscopic level, as follows.
First, the knowledge of the joint distribution of all
$\xi_{\vec r, \muv}(t)$ allows to write the joint
distribution of the full history of $\eta$ as
\begin{align}
  P[\eta,\vec{j}] &=
  \left\langle
  \prod_{\vec{r},t}
  \delta_{\eta_{\vec{r}}(t+\dd t) - \eta_{\vec{r}}(t),
   \dd t \sum_{\nuv} \left(\vec{j}_{\rv-\nuv}(t) -\vec{j}_{\rv}(t) \right) \cdot \nuv}
  \times
  \delta_{\vec{j}_{\rv}(t) \dd t,
    \sum_{\nuv} \nuv [\eta_{\vec{r}}(1-\eta_{\vec{r}+\nuv})  \xi_{\vec{r},\vec{\nu}}(t)
  - \eta_{\vec{r}+\nuv}(1-\eta_{\vec{r}})  \xi_{\vec{r}+\vec{\nu},-\vec{\nu}}(t) ]}
  \right\rangle_{\xi} \n \\
  &=\prod_{\vec{r},t}
   \delta_{\eta_{\vec{r}}(t+\dd t) - \eta_{\vec{r}}(t),
   \dd t \sum_{\nuv} \left(\vec{j}_{\rv-\nuv}(t) -\vec{j}_{\rv}(t) \right) \cdot \nuv}
  \times
  \moy{
    \delta_{\vec{j}_{\rv}(t) \dd t,
    \sum_{\nuv} \nuv [\eta_{\vec{r}}(1-\eta_{\vec{r}+\nuv})  \xi_{\vec{r},\vec{\nu}}(t)
  - \eta_{\vec{r}+\nuv}(1-\eta_{\vec{r}})  \xi_{\vec{r}+\vec{\nu},-\vec{\nu}}(t) ]}
  }_{\xi}
  \:,
    \label{eq:DefProbJointDiscrMFT}
\end{align}
where in the second line we used the fact that all the variables $\xi_{\muv,t}$ are independent.
We then rewrite each Kroenecker delta using its integral representation as
\begin{align}
  \delta_{\eta_{\vec{r}}(t+\dd t) - \eta_{\vec{r}}(t), \dd t \sum_{\nuv} \left(\vec{j}_{\rv-\nuv}(t) -\vec{j}_{\rv}(t) \right) \cdot \nuv}
  = \int_{-\pi}^\pi \frac{\dd \theta_{r}(t)}{2\pi} \exp \left[
    \I \theta_r(t) \left(
      \eta_{\vec{r}}(t+\dd t) - \eta_{\vec{r}}(t)
      -\dd t \sum_{\nuv} \left(\vec{j}_{\rv-\nuv}(t) -\vec{j}_{\rv}(t) \right) \cdot \nuv
    \right)
  \right]
  \:,
  \label{eq:delta_first}
\end{align}
and similarly
\begin{align}
  \label{eq:IntegRepresKronQ}
  &\delta_{\vec{j}_{\rv}(t) \dd t,
    \sum_{\nuv} \nuv [\eta_{\vec{r}}(1-\eta_{\vec{r}+\nuv})  \xi_{\vec{r},\vec{\nu}}(t)
    - \eta_{\vec{r}+\nuv}(1-\eta_{\vec{r}})  \xi_{\vec{r}+\vec{\nu},-\vec{\nu}}(t) ]} \n\\
  &= \int_{-\pi}^\pi \frac{\dd \vec{\varphi}_{\vec r}(t)}{(2\pi)^D} \exp \left[
  \I \vec{\varphi}_{\vec r}(t) \cdot \left(\vec{j}_{\rv}(t) \dd t
  \sum_{\nuv} \nuv [\eta_{\vec{r}}(1-\eta_{\vec{r}+\nuv})  \xi_{\vec{r},\vec{\nu}}(t)
  - \eta_{\vec{r}+\nuv}(1-\eta_{\vec{r}})  \xi_{\vec{r}+\vec{\nu},-\vec{\nu}}(t)]  \right)
  \right]
  \:.
\end{align}
We can now easily compute the average over $\xi$ that appears in~\cref{eq:DefProbJointDiscrMFT} by using its
distribution given in \cref{eq:multi_xi}, finding
\begin{align}
  &\moy{ \delta_{\vec{j}_{\rv}(t) \dd t,
    \sum_{\nuv} \nuv [\eta_{\vec{r}}(1-\eta_{\vec{r}+\nuv})  \xi_{\vec{r},\vec{\nu}}(t)
    - \eta_{\vec{r}+\nuv}(1-\eta_{\vec{r}})  \xi_{\vec{r}+\vec{\nu},-\vec{\nu}}(t) ]
    } }_{\xi}
  \n\\
  &= \int_{-\pi}^\pi \frac{\dd \vec{\varphi}_{\vec r}(t)}{(2\pi)^D}\,
  e^{ \I \vec{\varphi}_{\vec r}(t) \cdot \vec{j}_{\rv}(t) \dd t}
 \prod_{\nuv}
  \moy{\e^{- \I \vec{\varphi}_{\vec r}(t) \cdot \nuv \eta_{\vec{r}}(1-\eta_{\vec{r}+\nuv}) \xi_{\vec{r},\vec{\nu}}(t)}
   \e^{\I \vec{\varphi}_{\vec r}(t) \cdot \nuv  \eta_{\vec{r}+\nuv}(1-\eta_{\vec{r}})   \xi_{\vec{r}+\vec{\nu},-\vec{\nu}}(t)} }_{\xi}
  \n\\
  &= \int_{-\pi}^\pi \frac{\dd \vec{\varphi}_{\vec r}(t)}{(2\pi)^D}\,
  e^{ \I \vec{\varphi}_{\vec r}(t) \cdot  \vec{j}_{\rv}(t) \dd t}
   \prod_{\nuv}
  \left[ 1 +  \eta_{\vec{r}}(1-\eta_{\vec{r}+\nuv})
    \gamma \dd t \left( \e^{- \I \vec{\varphi}_{\vec r}(t) \cdot \nuv } - 1 \right)  \right]
  \left[ 1 +  \eta_{\vec{r}+\nuv}(1-\eta_{\vec{r}})
    \gamma \dd t \left( \e^{\I \vec{\varphi}_{\vec r}(t) \cdot \nuv} - 1 \right)  \right]
  \:.
\end{align}
Up to first order in $\dd t$ we can exponentiate
\begin{align}
  &\moy{ \delta_{\vec{j}_{\rv}(t) \dd t,
    \sum_{\nuv} \nuv [\eta_{\vec{r}}(1-\eta_{\vec{r}+\nuv})  \xi_{\vec{r},\vec{\nu}}(t)
    - \eta_{\vec{r}+\nuv}(1-\eta_{\vec{r}})  \xi_{\vec{r}+\vec{\nu},-\vec{\nu}}(t) ]} }_{\xi}
  \n\\
  &= \int_{-\pi}^\pi \frac{\dd \vec{\varphi}_{\vec r}(t)}{(2\pi)^D}
  \exp{  \I \vec{\varphi}_{\vec r}(t) \cdot  \vec{j}_{\rv}(t) \dd t
  + \gamma \dd t \sum_{\nuv}
  \left[ \eta_{\vec{r}}(1-\eta_{\vec{r}+\nuv}) (\e^{- \I \vec{\varphi}_{\vec r}(t) \cdot \nuv}-1)
  + \eta_{\vec{r}+\nuv}(1-\eta_{\vec{r}}) (\e^{\I \vec{\varphi}_{\vec r}(t) \cdot \nuv} - 1) \right]}
  \:,
\end{align}
which together with \cref{eq:delta_first} allows to rewrite the
probability distribution in \cref{eq:DefProbJointDiscrMFT} as
\begin{align}
  P[\eta,\vec{j}] =
  \int_{-\pi}^{\pi} \prod_{\vec{r},t} \frac{\dd \vec{\varphi}_{\vec r}(t)}{(2\pi)^D}
  \frac{\dd \theta_{\vec{r}}(t)}{2\pi}
  \exp \Bigg\lbrace &
  \sum_{t=0}^T \sum_{\vec{r}}  \Bigg[  \I \vec{\varphi}_{\vec r}(t) \cdot \vec{j}_{\rv}(t) \dd t - 
  \I \theta_r(t) \left(
    \eta_{\vec{r}}(t+\dd t) - \eta_{\vec{r}}(t)
    -\dd t \sum_{\nuv} \left(\vec{j}_{\rv-\nuv}(t) -\vec{j}_{\rv}(t) \right) \cdot \nuv
  \right)
  \n \\
  &+\gamma \dd t \sum_{\nuv} 
  \Bigg( \eta_{\vec{r}}(1-\eta_{\vec{r}+\nuv}) (\e^{- \I \vec{\varphi}_{\vec r}(t) \cdot \nuv}-1)
  + \eta_{\vec{r}+\nuv}(1-\eta_{\vec{r}}) (\e^{\I \vec{\varphi}_{\vec r}(t) \cdot \nuv} - 1) \Bigg)
  \Bigg]
  \Bigg\rbrace
  \:.
\end{align}
Taking the limit $\dd t \to 0$ (i.e.~$M\to \infty$ with $T$ fixed) finally yields
\begin{align}
  P[\eta,\vec{j}] = \int_{-\pi}^{\pi} \prod_{\vec{r},t} \frac{\dd \vec{\varphi}_{\vec r}(t)}{(2\pi)^D}
  \frac{\dd \theta_{\vec{r}}(t)}{2\pi}
    \exp \Bigg\lbrace &
    \int_0^T \dd{t} \sum_{\vec{r}} \Bigg[  \I  \vec{\varphi}_{\vec r} \cdot \vec{j}_{\rv}(t)
    - \I \theta_r \left( \dt{}{t} \eta_{\vec{r}}
      -\sum_{\nuv} \left(\vec{j}_{\rv-\nuv}(t) -\vec{j}_{\rv}(t) \right) \cdot \nuv
    \right)
    \n\\
     &+\gamma \sum_{\nuv} 
    \Bigg( \eta_{\vec{r}}(1-\eta_{\vec{r}+\nuv}) (\e^{- \I \vec{\varphi}_{\vec r} \cdot \nuv}-1)
    + \eta_{\vec{r}+\nuv}(1-\eta_{\vec{r}}) (\e^{\I \vec{\varphi}_{\vec r} \cdot \nuv} - 1) \Bigg)
    \Bigg]
    \Bigg\rbrace
    \:.  \label{eq:DiscrMFT}
\end{align}

In the following, it will prove convenient to Wick-rotate the two Lagrange multipliers as $\mathrm{i}\theta \mapsto \theta$ and $\mathrm{i}\vec \varphi\mapsto \vec \varphi$, which gives the real action
\begin{align}
    \cor S &= \int_0^T \dd{t} \cor L =\int_0^T \dd{t} \sum\r \cor L\r[\{\eta,\vec j,\theta,\vec \varphi \}], \label{eq:action1} \\
    \cor L\r &\equiv -\vec \varphi\r \cdot  \vec j\r
    +\theta\r \left[ \dot \eta\r - \sum_{\vec \nu} \left( \vec j\rmn-\vec j\r \right)\cdot \vec \nu \right] 
    -\gamma \sum_{\vec \nu} \left[\eta\r \left(1-\eta\rpn \right)\left(e^{- \vec \nu\cdot \vec \varphi\r}-1\right)+\eta\rpn \left(1-\eta\r\right)\left(e^{ \vec\nu\cdot \vec\varphi\r}-1\right)\right], \label{eq:action2}
\end{align}
where again the sum over $\vec r$ runs over all the sites of a $D$-dimensional cubic lattice, while $\vec \nu \in \{ \vec{e}_1, \vec e_2,\dots ,\vec e_D \}$. Note that in this final expression we have omitted everywhere the time dependence, so as to lighten the notation. The joint probability distribution of the occupations $\eta\r(t)$ and currents $\vec j\r(t)$ then reads
\begin{equation}
    P[\eta, \vec j]=\int_{-\I \pi}^{\I\pi} \prod_{\vec{r},t} \frac{\dd \vec{\varphi}_{\vec r}(t)}{(2\pi)^D}
  \frac{\dd \theta_{\vec{r}}(t)}{2\pi} \, e^{-\cor S}.
    \label{eq:joint_prob}
\end{equation}

\subsubsection{Application to the integrated current}
\label{sec:SP}

To tackle the current fluctuations at the origin $\vec r=\vec 0$, we first introduce the generating functional
\begin{equation}
    \cor Z[\vec \lambda] \equiv \expval{  \exp( 
    \vec \lambda
    \cdot \int_0^T \dd{t} \vec j_{\vec 0}(t)) },
    \label{eq:genf}
\end{equation}
where the average is intended over $P[\eta, \vec j]$ given in \cref{eq:joint_prob}.
We then minimize the quantity $[\cor S - \vec \lambda \cdot \int_0^T \dd{t} \vec j_{\vec 0}(t) ]$ with respect to the fields $\eta\r(t)$, $\vec j\r(t)$, and to the Lagrange multipliers $\theta\r(t)$, $\vec \varphi\r(t)$, as detailed in \cref{app:detail_minimization}.
Such minimizing procedure renders the following saddle-point equations.

\medskip\noindent
\textit{Occupations:} grouping all terms proportional to $\delta \theta\r(t)$, we find
\begin{equation}
        \dot \eta\r = \sum_{\vec \nu} \vec \nu \cdot \left( \vec j\rmn-\vec j\r \right).
    \label{eq:eta}
\end{equation}
\textit{Currents:} from the terms proportional to $\delta \vec \varphi\r(t)$,
\begin{equation}
        \left(\vec j\r\right)_\nu = \gamma\left[ \eta\r(1-\eta\rpn)e^{-\vec \nu \cdot \vec \varphi\r} -\eta\rpn(1-\eta\r)e^{\vec\nu\cdot \vec \varphi\r} 
        \right]. 
    \label{eq:j}
\end{equation}
\textit{Lagrange multipliers:} from the terms proportional to $\delta \vec j\r(t)$ and $\delta \eta\r(t)$, respectively,
\begin{align}
        &\left(\vec \varphi\r\right)_\nu = \theta\r-\theta\rpn - \delta_{\vec r,\vec 0}\lambda_\nu, \label{eq:varphi} \\
        &\;\, \dot \theta\r = -\gamma \sum_{\vec \nu} \left[
        \left(1-\eta\rpn \right)\left(e^{- \vec \nu\cdot \vec \varphi\r}-1\right)-\eta\rpn \left(e^{ \vec\nu\cdot \vec\varphi\r}-1\right)
        +\left(1-\eta\rmn \right)\left(e^{\vec \nu\cdot \vec \varphi\rmn}-1\right)-\eta\rmn \left(e^{ -\vec\nu\cdot \vec\varphi\rmn}-1\right)
        \right],    \label{eq:theta}
\end{align}
with the constraint
\begin{equation}
    \theta\r(t=T)=0.
    \label{eq:final_condition}
\end{equation}
In all these equations, the subscript $\nu$ indicates the $\nu-$th component of a given (vectorial) quantity. In the following, we will choose without loss of generality $\vec \lambda = \lambda \, \vec e_1$, and study the fluctuations of the first component of the current $\vec j_{\vec 0}$.

Crucially, note that by taking variations as $\delta \eta\r(t)$ (see \cref{app:detail_minimization}) we are seemingly disregarding the discrete nature of the occupations $\eta\r(t)$. In principle, this is automatically taken into account by having enforced the conservation relations~\eqref{eq:multi_eta}~to~\eqref{eq:multi_xi},
for a suitable choice of
the initial conditions~\cite{Lefevre:2007}. 
Still, we envision that a more careful treatment may be necessary in order to correctly capture the correlations between neighbouring sites, which is especially relevant in finite 1D systems --- see the discussion in \cref{app:micro_1D}. Since these correlations vanish in infinite systems at equilibrium (see e.g.~\cref{sec:MasterEq2D}), this point is however immaterial for what concerns the following discussion, which is restricted to regular lattices in $D\geq 2$.

Note that, at this stage, all the fields that appear in \cref{eq:eta,eq:j,eq:theta,eq:varphi} are to be considered time dependent. If a stationary solution exists, however, one expects to be able to write at long times $T$
\begin{equation}
    \cor S = \int_0^T \dd t \cor L(t) \simeq T \cor L^*,
    \label{eq:assumptions}
\end{equation}
where $\cor L^*$ is the Lagrangian of \cref{eq:action1,eq:action2} computed in correspondence of the stationary fields.
Focusing on the stationary solution has been sometimes justified \textit{a priori}, in $D=1$ (and unless dynamical phase transitions are expected~\cite{Bertini:2015}), on the basis of an \textit{additivity principle} conjectured in \cite{Bodineau:2004}. 
In the following, however, we will not assume time independence, but we will rather determine that the time-independent solution of \cref{eq:eta,eq:j,eq:theta,eq:varphi} indeed provides the dominant contribution at long times.

Still, as they stand, \cref{eq:eta,eq:j,eq:theta,eq:varphi} cannot be readily solved because they are nonlinear in $\vec \varphi\r$. To make progress, one can expand all the fields in powers of $\lambda$, e.g. 
\begin{equation}
    \eta\r = \eta\z\r+\lambda \eta\r\o + \lambda^2 \eta\r\t +\dots,
\end{equation}
and collect terms proportional to equal powers of $\lambda$. Note, however, that the saddle-point equations would remain nonlinear unless the leading-order term $\vec\varphi\z\r $ vanishes identically.
Under the assumption that $\vec\varphi\z\r \equiv  \vec 0$, which we adopt henceforth, the saddle-point equations become linear and can in principle be solved to any desired order in $\lambda$, thus giving access to the various cumulants. We will justify this assumption \textit{a posteriori}, by comparing our final result both to numerical simulations, and to the fluctuations of the integrated current predicted in \cref{eq:FlucCurrent2D} starting from the master equation, finding in both cases an excellent agreement.

\medskip
\textit{Solution at $\order{\lambda^0}$}. ---
We first search for a solution of the saddle-point \cref{eq:eta,eq:j,eq:theta,eq:varphi} with $\vec\varphi\z\r \equiv  \vec 0$, up to their leading order in $\lambda$. By inspecting \cref{eq:theta}, which rules the evolution of $\theta\r(t)$, the assumption above implies
\begin{equation}
    \dot \theta\r\z = 0,
\end{equation}
while \cref{eq:varphi} gives
\begin{equation}
    \theta\r\z = \theta\rpn\z.
\end{equation}
Using the final condition $\theta\r(t=T)=0$ in \cref{eq:final_condition}, we then deduce that $\theta\r\z(t) \equiv 0$ identically for all $\vec r$ and $t$.
There remains to solve \cref{eq:j,eq:eta} for the occupations and the currents, which simplify as
\begin{equation}
    \dot \eta\r\z = \sum_{\vec \nu} \vec \nu \cdot \left( \vec j\rmn\z-\vec j\r\z \right),
    \label{eq:eta_0}
\end{equation}
and
\begin{equation}
        \left(\vec j\r\z\right)_\nu  =\gamma \left( \eta\r\z -\eta\rpn\z\right),
    \label{eq:j_0}
\end{equation}
respectively. Note that $\vec j\r\z$ can be easily eliminated from \cref{eq:eta_0} by using \cref{eq:j_0}, which gives
\begin{equation}
    \partial_t \eta\r\z = \sum_{\vec \nu}\left( 
    \eta\rpn\z+\eta\rmn\z  -2\eta\r\z\right) \equiv \Delta \eta\r\z,
\end{equation}
where in the last step we recognized the discrete Laplace operator. 
This equation being dissipative, its solutions decay toward a spatially homogeneous constant after an initial temporal transient.
At long times, the contribution of this transient solution is not extensive in $T$, which is why we discard it and focus instead on the stationary solution. Here
\begin{equation}
    \eta\r\z \equiv \rb,
    \label{eq:occupations_zero}
\end{equation}
where we identified the constant value assumed everywhere by $\eta\r\z$ as the average density $\rb$ of the system, and finally from \cref{eq:j_0} we obtain trivially
\begin{equation}
    \vec j\r\z = \vec 0.
    \label{eq:currents_solution_0}
\end{equation}

\medskip
\textit{Solution at $\order{\lambda}$}. ---
We now collect terms of $\order{\lambda}$ in the saddle-point \cref{eq:eta,eq:j,eq:theta,eq:varphi}, assume again $\vec\varphi\z\r\equiv \vec 0$, and insert $\eta\r\z \equiv \rb$ from \cref{eq:occupations_zero}. This renders the following relations for the occupations,
\begin{equation}
    \dot \eta\r\o = \sum_{\vec \nu} \vec \nu \cdot \left( \vec j\rmn\o-\vec j\r\o \right),
    \label{eq:eta_1}
\end{equation}
for the currents,
\begin{equation}
    \left(\vec j\r\o\right)_\nu  =\gamma \left[ \eta\r\o -\eta\rpn\o
    -2\rb(1-\rb) \left(\vec \varphi\r\o\right)_\nu
    \right],
    \label{eq:j_1}    
\end{equation}
and finally for the Lagrange multipliers,
\begin{align}
        &\left(\vec \varphi\r\o\right)_\nu = \theta\r\o-\theta\rpn\o - \delta_{\vec r,\vec 0}\,\delta_{\nu,1}, \label{eq:varphi_1} \\
        &\;\, \dot \theta\r = -\gamma \sum_{\vec \nu} \vec \nu \cdot \left( \vec\varphi \rmn\o-\vec\varphi\r\o \right),
        \label{eq:theta_1}
\end{align}
with the constraint $\theta\r\o(t=T)=0$. 

We note from the onset that $\vec\varphi\r\o=0$ is no longer a valid solution. However, $\vec\varphi\r\o$ can be promptly eliminated from \cref{eq:theta_1} by using \cref{eq:varphi_1}, which gives
\begin{align}
    \partial_t \theta\r\o &= -\gamma \sum_{\vec \nu}\left( 
    \theta\rpn\o+\theta\rmn\o  -2\theta\r\o\right) +\gamma \left( \delta_{\vec r,\vec e_1}-\delta_{\vec r,\vec 0}  \right) \n\\
    &= -\gamma \Delta \theta\r\o +\gamma \left( \delta_{\vec r,\vec e_1}-\delta_{\vec r,\vec 0}  \right) .
    \label{eq:theta_1_alone}
\end{align}
This anti-diffusion equation can in principle be solved for any time $t$ by introducing the inverted time variable $\tau=T-t$, so that $\theta\o\r(t=T)=0$ becomes an initial condition. However, note that \cref{eq:theta_1_alone} this way becomes a \textit{bona fide} discrete diffusion equation in $\tau$, hence we expect it to eventually reach a stationary solution for sufficiently large $T$. This stationary solution dominates the saddle-point integral, in agreement with the expectations spelled out under \cref{eq:assumptions}.
We are thus left with
\begin{equation}
    \Delta \theta\r\o = \left( \delta_{\vec r,\vec e_1}-\delta_{\vec r,\vec 0}  \right) ,
    \label{eq:theta_1_stat}
\end{equation}
which can be easily solved in Fourier space by introducing
\begin{equation}
    \theta\r\o = \int_\T{BZ} \dslash{q} e^{-i\vec q\cdot \vec r} \theta\q\o \qquad \leftrightarrow \qquad \theta\q\o = \sum\r e^{i\vec q\cdot \vec r}\theta\r\o,
\end{equation}
where each of the $D$ integrals is intended over the Brillouin zone $q_j\in[-\pi,\pi]$. Fourier transforming \cref{eq:theta_1_stat} gives
\begin{equation}
    2 \, \theta\q\o \sum_{\nu=1}^D \left[ \cos(q_\nu)-1 \right] = e^{i q_1}-1,
\end{equation}
and thus
\begin{equation}
    \theta\r\o = \int_\T{BZ} \dslash{q} e^{-i\vec q\cdot \vec r} \frac{e^{i q_1}-1}{2 \sum_{\nu=1}^D \left[ \cos(q_\nu)-1 \right]}.
    \label{eq:theta1_sol}
\end{equation}
One can check that this function is real, and has a maximum at $\vec{r}=\vec 0$, where $\theta\o_{\vec 0} =\frac{1}{2D}$. Note that the defining integral~\eqref{eq:joint_prob} for $P[\eta, \vec j]$ was intended along the imaginary axis, while the stationary field $\theta\r$ found in \cref{eq:theta1_sol} is real;
this requires a deformation of the complex integration contour that we assume to be legit, the integrand function $\exp(-\cor S)$ being analytic.

The stationary solution for $\vec \varphi\r\o$ is then found immediately, using its saddle-point equation~\eqref{eq:varphi_1},
in terms of $\theta\r\o$ given in \cref{eq:theta1_sol}. This can be used to reconstruct the stationary currents $\vec j\o\r$ given in \cref{eq:j_1}, provided that $\eta\r\o$ is known first. We then focus on $\eta\r\o$, which can be found from its saddle-point equation~\eqref{eq:eta_1} by eliminating $\vec j\r\o$ using \cref{eq:j_1}. This gives
\begin{align}
    \dot \eta\r\o &= \gamma \Delta \eta\r\o
    +2\rb(1-\rb)\gamma \sum_{\vec \nu} \vec \nu \cdot \left( \vec \varphi\r\o-\vec \varphi\rmn\o \right) \n\\
    &=    \gamma \Delta \eta\r\o
    +2\rb(1-\rb)\gamma \left[ -\Delta \theta\r\o  +\left( \delta_{\vec r,\vec e_1}-\delta_{\vec r,\vec 0}  \right) \right],
    \label{eq:eta1_step}
\end{align}
where in the second step we replaced $\varphi\r\o$ using \cref{eq:varphi_1}. As we commented above, the dominant contribution comes from enforcing $\partial_t \theta\r\o=0$, and hence the stationarity condition for $\theta\r\o$ in \cref{eq:theta_1_stat} simplifies \cref{eq:eta1_step} as
\begin{equation}
    \dot \eta\r\o =    \gamma \Delta \eta\r\o.
\end{equation}
This simple diffusion equation for $\eta\r\o$ admits a spatially uniform solution at long times. To find it, we note that at large distance it must be $\eta\r \to \rb$, and thus in particular $\eta\r\o \to 0$. This fixes the stationary solution
\begin{equation}
    \label{eq:Eta2DOrder1}
    \eta\r\o \equiv 0.
\end{equation}

We can finally use \cref{eq:j_1} to find the currents $\vec j\r\o$ as
\begin{align}
    \left(\vec j\r\o\right)_\nu  &=-2\gamma \rb(1-\rb) \left(\vec \varphi\r\o\right)_\nu \n\\
    &= -2\gamma \rb(1-\rb)\left[\theta\r\o-\theta\rpn\o - \delta_{\vec r,\vec 0}\,\delta_{\nu,1}\right] \n\\
    &= 2\gamma \rb(1-\rb) \left[\delta_{\vec r,\vec 0}\,\delta_{\nu,1} + \int_\T{BZ} \dslash{q} e^{-i\vec q\cdot \vec r} \frac{(e^{-i q_\nu}-1)(e^{i q_1}-1)}{2 \sum_{\mu=1}^D \left[ \cos(q_\mu)-1 \right]}   \right],
    \label{eq:currents_solution}    
\end{align}
where in the second and third step we used \cref{eq:varphi_1,eq:theta1_sol}, respectively.

\medskip
\textit{Solution for the fluctuations of the integrated current}. ---
Equation~\eqref{eq:currents_solution} can be used to estimate the fluctuations of the integrated current $Q_T$ through the bond $(0,0)$--$(0,1)$, when $T$ is large.
Indeed, in \cref{eq:genf} we introduced its generating functional
\begin{equation}
    \cor Z[\vec \lambda] = \int \cor D [\eta, \vec j, \theta, \vec \varphi]\, \exp[ -\cor S+
    \vec \lambda
    \cdot \int_0^T \dd{t} \vec j_{\vec 0}(t)]  \propto \exp[ -\cor S^*+
    \vec \lambda
    \cdot \int_0^T \dd{t} \vec j_{\vec 0}^*(t)] ,
\end{equation}
where we indicated by $\cor S^*$ the action in \cref{eq:action1} computed in correspondence of the saddle-point fields, and similarly $\vec j_{\vec 0}^*(t)$ is the saddle-point current. In our case $\vec \lambda = \lambda \vec e_1$ and the saddle-point current is stationary, whence
\begin{equation}
    \cor Z[\lambda] \propto \exp[ -\cor S^*+
    \lambda
    T \left(\vec j_{\vec 0}^*\right)_1], 
\end{equation}
where $\vec j_{\vec 0}^*= \vec j_{\vec 0}\z+\lambda \vec j_{\vec 0}\o+\order{\lambda^2}$ is the solution we evaluated in this Section, see \cref{eq:currents_solution_0,eq:currents_solution}. Using that the action $\cor S$ is stationary by construction at the saddle point, we can finally compute
\begin{equation}
    \moy{Q_T^2} = \frac12 \dv[2]{\log \cor Z[\lambda]}{\lambda} \simeq T \left( \vec j_{\vec r = \vec 0}\o \right)_1 = 2\gamma \rb(1-\rb) \left(1-\frac{1}{D}\right) T,
    \label{eq:fluc_current_d}
\end{equation}
where 
\rev{we have identified}
$Q_T = \int_0^T \dd{t} \left( \vec j_{\vec r = \vec 0} (t) \right)_1$.
This coincides, upon setting $D=2$ and $\gamma=1$, with the exact result found in \cref{eq:FlucCurrent2D} using the master equation. Together with the results of the numerical simulations discussed in the next Section, this puts the assumptions underlying the derivation presented in this Section on firmer grounds. Besides, the microscopic approach presented here allows in principle to access higher-order moments of $Q_T$ simply by refining the solution for $\vec j\r$ toward higher orders in $\lambda$.

\subsubsection{Discussion and current-current correlation profile}
\label{sec:DiscCurrProf}

We now discuss the physical interpretation of the quantities computed in \cref{sec:SP} from the microscopic action formalism. Following the ideas developed in \cref{sec:int_current_comb_mft} in the MFT formalism, we can show that
\begin{equation}
    \frac{\expval{\eta\r(t) \, e^{\lambda Q_T}}}{\expval{e^{\lambda Q_T}}}
    =
    \frac{ \displaystyle
    \int \cor D [\eta, \vec j, \theta, \vec \varphi]\,
    \eta\r(t)
    \,\exp[ -\cor S+
    \vec \lambda
    \cdot \int_0^T \dd{t} \vec j_{\vec 0}(t)]
    }{ \displaystyle
    \int \cor D [\eta, \vec j, \theta, \vec \varphi]
    \,\exp[ -\cor S+
    \vec \lambda
    \cdot \int_0^T \dd{t} \vec j_{\vec 0}(t)]
    }
    \underset{T \to \infty}{\simeq} \eta\r^*(t)
    \:,
\end{equation}
because the saddle point is identical in the numerator and the denominator. In particular, at first order in $\lambda$, this yields
\begin{equation}
    \moy{\eta\r(t) Q_T} = \eta\r\o = 0
    \:,
\end{equation}
from~\eqref{eq:Eta2DOrder1}. This result is apparently in net contrast with the correlations $\moy{\eta\r(T) Q_T}$ computed in \cref{sec:MasterEq2D}, which do not vanish. But in fact, there is no contradiction since $\moy{\eta\r(T) Q_T} = \mathrm{Cov}(\eta\r(T), Q_T)$ measures the correlations between the integrated current and the occupation at the \textit{same time} $T$, while $\moy{\eta\r(t) Q_T} = \mathrm{Cov}(\eta\r(t), Q_T)$ measures these correlations at \textit{different times} $1 \ll t \ll T$. Note that both are stationary: $\moy{\eta\r(T) Q_T}$ does not depend on $T$ for $T \to \infty$; while, for a given large $T$, $\moy{\eta\r(t) Q_T}$ is independent of $t$.
A similar discussion can be found in~\cite{Derrida:2019,Derrida:2019b}.

\begin{figure}
    \centering
     \includegraphics[width=0.6\textwidth]{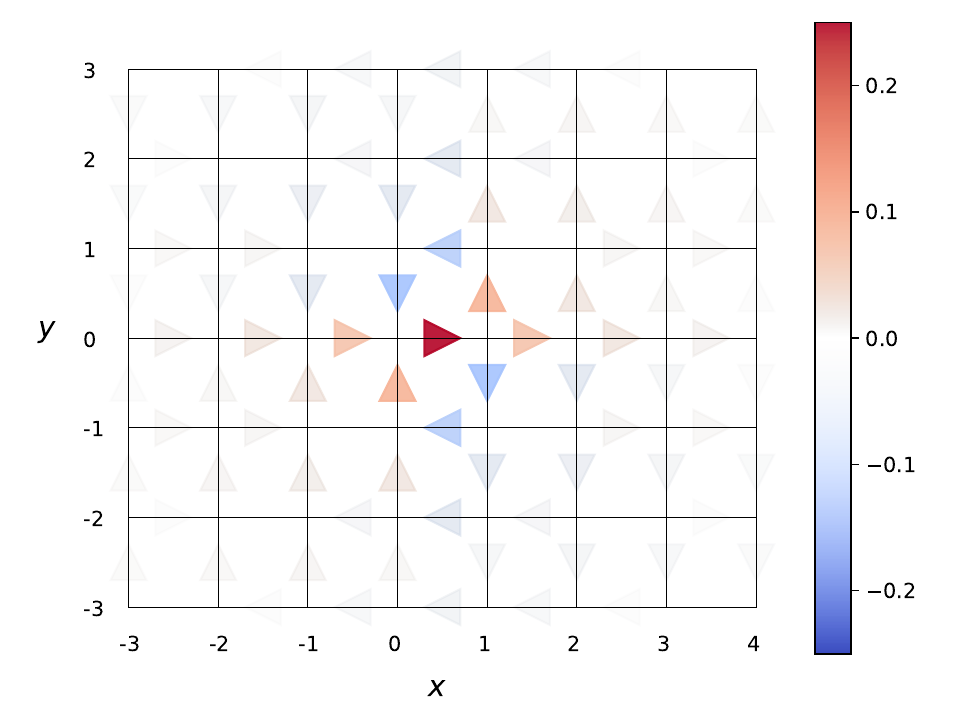}
    \caption{Plot of the solution $\vec j_{\vec r}\o$ given in \cref{eq:currents_solution} for $D=2$, corresponding to a rare realization of the dynamics in which a given particle circulates repeatedly through the horizontal bond at $\vec r=\vec 0$, thus producing two specular vortices. 
    In the plot $\vec j_{\vec r}\o$ is rescaled by the prefactor $2\gamma\rb(1-\rb)$ that appears in \cref{eq:currents_solution}.}
    \label{fig:currents}
\end{figure}

These results stress the importance of the choice of correlations to consider: $\moy{\eta\r(T) Q_T}$ contains enough information to compute $\moy{Q_T^2}$ (see \cref{sec:MasterEq2D}), but $\moy{\eta\r(t) Q_T}$ does not, since it vanishes. As we have seen above, in the stationary regime $1 \ll t \ll T$, the relevant information is encoded in the saddle-point currents $\vec j\r^*(t)$, from which $\moy{Q_T^2}$ is given in \cref{eq:fluc_current_d}. These quantities have the meaning of a current-current correlation profile, similarly given by
\begin{equation}
    \frac{\expval{\vec j\r(t) \, e^{\lambda Q_T}}}{\expval{e^{\lambda Q_T}}}
    =
    \frac{ \displaystyle
    \int \cor D [\eta, \vec j, \theta, \vec \varphi]\,
    \vec j\r(t)
    \,\exp[ -\cor S+
    \vec \lambda
    \cdot \int_0^T \dd{t} \vec j_{\vec 0}(t)]
    }{ \displaystyle
    \int \cor D [\eta, \vec j, \theta, \vec \varphi]
    \,\exp[ -\cor S+
    \vec \lambda
    \cdot \int_0^T \dd{t} \vec j_{\vec 0}(t)]
    }
    \underset{T \to \infty}{\simeq} \vec j\r^*(t)
    \:.
\end{equation}
Expanding at first order in $\lambda$ yields that the saddle-point solution~\eqref{eq:currents_solution} corresponds to
\begin{equation}
    \label{eq:PhysInterpJSP}
    \vec j\r^{(1)}(t) = \moy{\vec j\r(t) \, Q_T } - \moy{\vec j\r(t)}\moy{ Q_T}
    \equiv \mathrm{Cov}(\vec j\r(t), Q_T)
    \:.
\end{equation}
This current-current correlation profile
is shown in \cref{fig:currents} for the case $D=2$. 
Interestingly, it shows that an increase of $Q_T$ is correlated with a dynamics in which particles circulate repeatedly through the bond $(0,0)-(1,0)$. This induces two specular vortices around this bond.
For the SEP in dimension $D>1$, vortices have indeed been previously indicated to play a major role in determining the current fluctuations~\cite{Bodineau:2008}. 
In particular, in Ref.~\cite{Bodineau:2008} the particle current through an extended 2D slit \rev{of length $l$} was analyzed within the macroscopic framework of MFT \rev{for a large but finite system of size $L\gg l$}, and 
vortices
localized at the slit edges were shown to dominate the local current fluctuations \rev{at long times. However, these results being obtained from a macroscopic description, the vortices appeared in~\cite{Bodineau:2008} as singularities in the current field, leaving their microscopic structure out of reach. Here, thanks to a fully microscopic computation, we have characterised the microscopic structure of the vortex configuration that fully determines the long-time current fluctuations through a \textit{single} bond. We expect that a similar microscopic structure is present at the edges of an extended slit, even if the length of the slit is macroscopic.}
\rev{Crucially, the} existence of these \rev{vortex} configurations is permitted by the looped structure of the lattice, which thus plays a central role in the existence of a \rev{stationary} state for the correlations in infinite systems.

\begin{figure}
    \centering
    \includegraphics[width=0.45\textwidth]{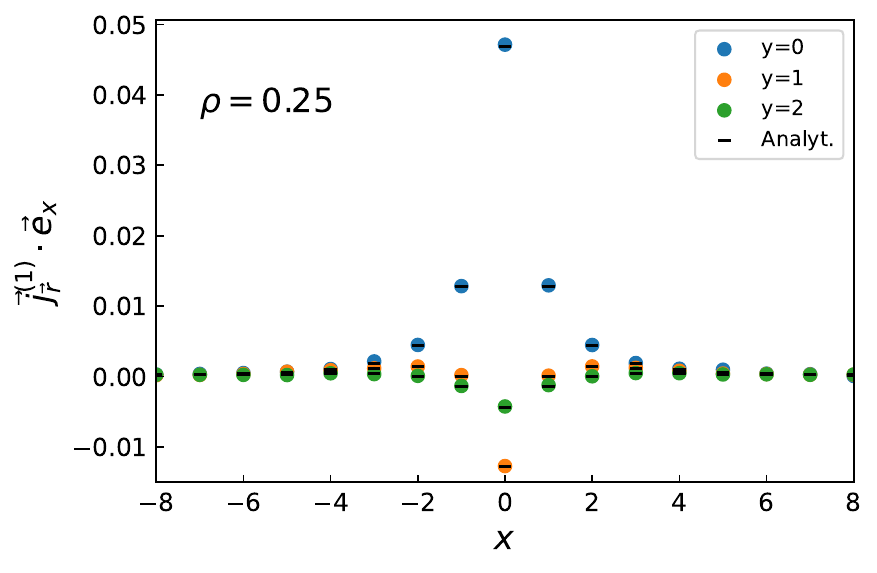}
    \includegraphics[width=0.45\textwidth]{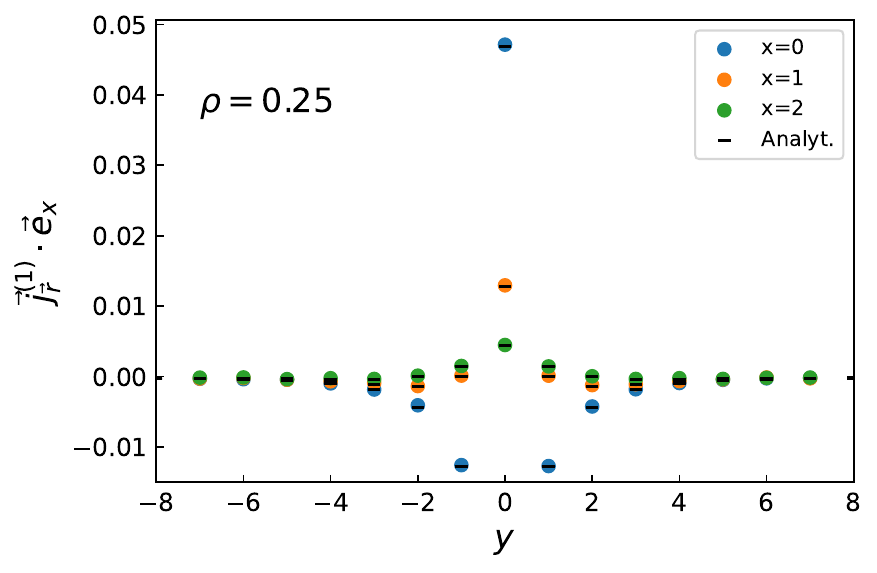}
    \includegraphics[width=0.45\textwidth]{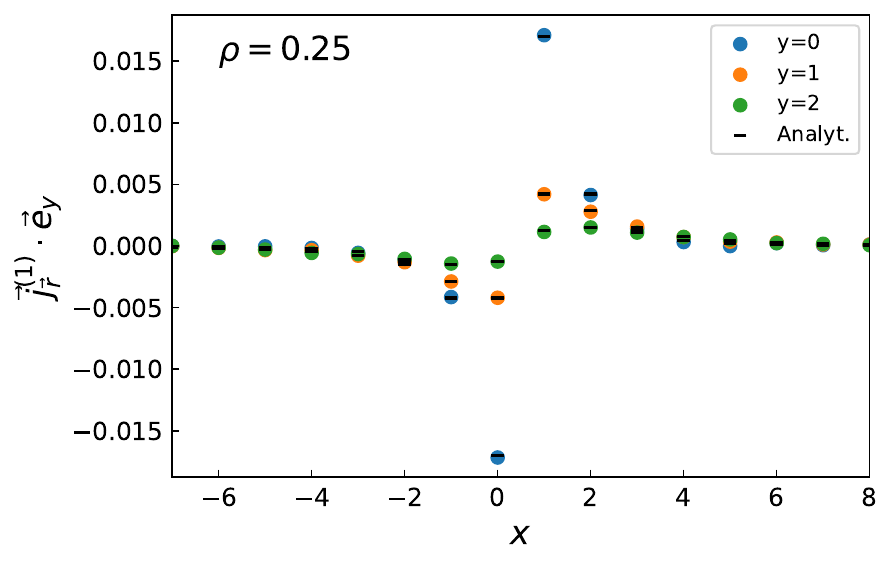}
    \includegraphics[width=0.45\textwidth]{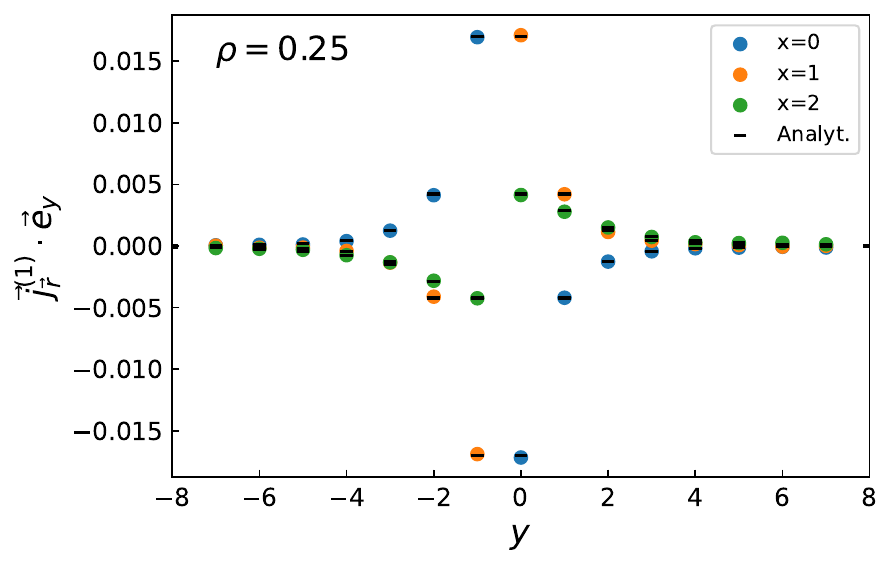}
    \caption{Current-current correlation profile $\vec{j}_{\vec{r}}^{(1)} = \langle \vec{j}_{\vec{r}}(t) \: Q_T \rangle$ in $d=2$, projected along the two components $\vec{e}_x$ and $\vec{e}_y$, as a function of $\vec{r} = (x,y)$. The \rev{dashes} correspond to the analytical solution~\eqref{eq:currents_solution}, while the \rev{points} are obtained from numerical simulations of the SEP in 2D. 
    Numerically, we computed these correlations as $ \frac{1}{T} \int_0^T \langle \vec{j}_{\vec{r}}(t) \: Q_T \rangle \dd t$ to improve the convergence by using the stationarity of $\vec{j}_{\vec{r}}^{(1)}$ for $1 \ll t \ll T$. The simulations are performed for a finite system of $100 \times 100$ sites with periodic boundary conditions, up to $T=500$, and averaged over $400000$ realisations.
    }
    \label{fig:Curr2DSim}
\end{figure}

Finally, relation~\eqref{eq:PhysInterpJSP} allowed us to compare our results~\eqref{eq:currents_solution} to numerical simulations performed on a 2D system. The comparison is shown in \cref{fig:Curr2DSim} and displays excellent agreement at various points on the lattice, and along both the horizontal and vertical directions.

\section{Conclusion}
\label{sec:Conclusion}

We have studied the fluctuations of the integrated current $Q_t$ through a given bond of the lattice, for the SEP on different infinite lattices. We have shown that on the comb the current-density correlation profiles $\moy{\eta_{\vec{r}}(t) Q_t}$ are not stationary (see \cref{eq:correlation_scaling_comb_Time}), leading to a sublinear growth of the integrated current fluctuations $\moy{Q_t^2}$ with time (see \cref{eq:FlucCurrentComb}). In contrast, we have shown that on hypercubic lattices in dimensions $D \geq 2$ the current-density correlation profiles are stationary and display a dipolar microscopic structure, leading to a linear scaling with time of $\moy{Q_t^2}$ --- see \cref{eq:CorrProf2D,eq:FlucCurrent2D} for the case $D=2$, and \cref{eq:fluc_current_d} for generic $D\geq 2$. We have identified 
the origin of these features 
in the looped structure of the lattice. Indeed, we have shown that the fluctuations of $Q_t$ are dominated by particles looping around the bond on which $Q_t$ is measured, thus leading to the emergence of vortices in the current-current correlation profile (see \cref{fig:currents}).

Our work opens several interesting directions for future investigation.
First, the MFT equations~\eqref{eq:MFTqComb}~to~\eqref{eq:BondCondMFTComb} derived here form a closed set that, if solved,
would give access to the full distribution of the integrated current $Q_t$ on the comb lattice. This formalism could also be applied to general diffusive systems on the comb beyond the SEP. Similarly, the microscopic action formalism we built in \cref{sec:micro_action} in principle provides a constructive method to compute arbitrarily high moments of $Q_t$, for regular lattices in dimension $D\geq 2$. Finally, singling out a tracer particle and studying the statistics of its displacement would allow to extend towards higher dimensions several interesting results that are already available in 1D~\cite{Krapivsky:2014,Krapivsky:2015a,Imamura:2017,Imamura:2021,Grabsch:2022}, or solely in the dense regime for other geometries~\cite{Brummelhuis:1988,Brummelhuis:1989,Benichou:2013,Benichou:2015,Illien:2018}.


\appendix
\numberwithin{equation}{section}

\section{Green's function for the diffusion equation on the comb}
\label{app:GreenFctComb}

In this Appendix, we aim to show that the expression~\eqref{eq:PropComb} is indeed the Green's function of the diffusion equation on the comb, i.e.~that it is solution of
\begin{equation}
    \label{eq:EqGreenComb}
    \partial_t G_t =  \delta(y) \partial_x^2 G_t + \partial_y^2 G_t
    \:,
    \quad \text{with} \quad
    G_{t=0}(x,y|x',y') = 
    \delta(x-x') \delta(y-y')
    \:.
\end{equation}
For this, we take the Laplace transform of~\eqref{eq:EqGreenComb}, which gives
\begin{equation}
    \label{eq:EqGreenCombLaplace}
    s \: \hat{G}_s(x,y|x',y') -  \delta(x-x') \delta(y-y')
    = [\delta(y) \partial_x^2 + \partial_y^2] \hat{G}_s(x,y|x',y'),
\end{equation}
where we have introduced
\begin{equation}
    \hat{G}_s(x,y|x',y') = \int_0^\infty \e^{-s t} G_t(x,y|x',y') \dd t
    \:.
\end{equation}
Starting from the expression of $\hat{G}_s$~\eqref{eq:PropComb},~i.e.
\begin{equation}
    \hat{G}_s(x,y|x',y') =
    \frac{1}{2 \sqrt{2} \: s^{1/4}}
    \e^{- \sqrt{2} s^{1/4} \abs{x-x'} - \sqrt{s}(\abs{y} + \abs{y'})}
    + \frac{1}{2 \sqrt{s}} \delta(x-x') \Theta(yy')
    \left(
    \e^{-\sqrt{s} \abs{y-y'}}
    - \e^{-\sqrt{s} \abs{y+y'}}
    \right)
    \:,
\end{equation}
we can compute the derivatives
\begin{multline}
    \partial_y^2 G_s(x,y|x',y')
    = \frac{1}{2\sqrt{2}} \e^{- \sqrt{2} s^{1/4} \abs{x-x'} - \sqrt{s} (\abs{y} + \abs{y'})}
  \left[
    s^{3/4} - 2 s^{1/4} \delta(y)
  \right]
  \\
  - \frac{1}{2} \delta(x-x') y'\delta(y y') \left(
    \sg{y-y'} \e^{-\sqrt{s} \abs{y-y'}}
    - \sg{y+y'} \e^{-\sqrt{s} \abs{y+y'}}
  \right)
  \\
  - \frac{1}{2} \delta(x-x') \Theta(yy') \left(
    2\delta(y-y')
    - 2\delta(y+y')
    - \sqrt{s} \: \e^{-\sqrt{s} \abs{y-y'}}
    + \sqrt{s} \: \e^{-\sqrt{s} \abs{y+y'}}
  \right)
  \:,
\end{multline}
and
\begin{multline}
  \partial_x^2 \hat{G}_s(x,y|x',y') =
  \frac{1}{2\sqrt{2} \: s^{1/4}}\e^{- \sqrt{2} s^{1/4} \abs{x-x'} - \sqrt{s} (\abs{y} + \abs{y'})}
  \left[
    2 \sqrt{s} - 2 \sqrt{2} s^{1/4} \delta(x-x')
  \right]
  \\
  + \frac{1}{2\sqrt{s}} \delta''(x-x') \Theta(yy')  \left(
    \e^{-\sqrt{s} \abs{y-y'}}
    -\e^{-\sqrt{s} \abs{y+y'}}
  \right)
  \:.
\end{multline}
Inserting these expressions into the r.h.s.~of~\eqref{eq:EqGreenCombLaplace}, and using that $y'\delta(y y') = \delta(y)/\sg{y'}$, we obtain that $\hat{G}_s$ is solution of~\eqref{eq:EqGreenCombLaplace}, and therefore that $G_t$ is the Green's function of the diffusion equation on the comb~\eqref{eq:EqGreenComb}.

A similar expression of $\hat{G}_s$ was obtained in~\cite{Grabsch:2023a} (although with different factors due to different jump rates considered in this article) by taking the continuous limit of the propagator of a random walk on the comb lattice computed in~\cite{Illien:2016}.

\bigskip

The Green's function $G_t$ as defined by Eq.~\eqref{eq:EqGreenComb} can be used to write the solution of the diffusion equation with an initial condition,
\begin{equation}
    \partial_t \rho =  \delta(y) \partial_x^2 \rho + \partial_y^2 \rho
    \:,
    \quad \text{with} \quad
    \rho(x,y;t=0) = \rho_0(x,y)
    \:,
\end{equation}
as
\begin{equation}
    \rho(x,y;t) = \int \dd x' \dd y' G_t(x,y|x',y') \rho_0(x',y')
    \:,
\end{equation}
by linearity of the equation. Importantly, it can also be used to obtain the solution of the equation with a source term,
\begin{equation}
    \label{eq:DiffEqCombSource}
    \partial_t \rho =  \delta(y) \partial_x^2 \rho + \partial_y^2 \rho + f
    \:,
    \quad \text{with} \quad
    \rho(x,y;t=0) = 0
    \:,
\end{equation}
since by defining $K_t = \Theta(t-0^+) G_t$, we can easily show that $K_t$ obeys
\begin{equation}
    \partial_t K_t(x,y|x',t') =  [\delta(y) \partial_x^2 + \partial_y^2] K_t(x,y|x',y')
    + \delta(t) \delta(x-x') \delta(y-y')
    \:.
\end{equation}
The solution of~\eqref{eq:DiffEqCombSource} can thus be written as
\begin{equation}
    \rho(x,y;t) = \int_0^t \dd t' \int \dd x'\dd y'
    G_{t-t'}(x,y|x',y') f(x',y';t)
    \:,
\end{equation}
again by linearity of the equation.

\section{Details of the minimization of the \texorpdfstring{$D$}{D}-dimensional microscopic action}
\label{app:detail_minimization}

To obtain the saddle-point equations reported in \cref{sec:SP},
we compute the variation of each individual term that appears in the dynamical action given in \cref{eq:action1,eq:action2}. For instance, the first term reads simply
\begin{equation}
    \delta \int_0^T \dd{t} \sum\r \vec \varphi\r(t) \cdot  \vec j\r(t) = \int_0^T \dd{t} \sum\r
    \left[ \vec \varphi\r(t) \cdot  \delta \vec j\r(t)+\vec j\r(t) \cdot \delta  \vec \varphi \r(t)  \right].
    \label{appeq:1}
\end{equation}
The second term gives, upon integrating by parts,
\begin{align}
    &\delta \int_0^T \dd{t} \sum\r \theta\r(t) \left[ \dot \eta\r(t) - \sum_{\vec \nu} \left( \vec j\rmn(t)-\vec j\r(t) \right)\cdot \vec \nu \right] \n\\
    &= \int_0^T \dd{t} \sum\r  \left\lbrace\left[ \dot \eta\r(t) - \sum_{\vec \nu} \left( \vec j\rmn(t)-\vec j\r(t) \right)\cdot \vec \nu \right]  \delta \theta\r(t) 
    +  \theta\r(t) \sum_{\vec \nu} \vec \nu \cdot \left[\delta \vec j\r(t) -\delta \vec j\rmn(t) \right] -\dot \theta\r(t) \delta \eta\r(t)
    \right\rbrace \n\\
    &\quad +\sum\r\left [ \theta\r(T)\delta \eta\r(T) -  \theta\r(0)\delta \eta\r(0) \right].
    \label{appeq:2}
\end{align}
The term $\delta \eta\r(0)$ requires some discussion: indeed, one may in principle consider either fixed or fluctuating initial occupations $\eta\r(0)$, corresponding respectively to a \textit{quenched} or \textit{annealed} initial condition (see e.g.~Ref.~\cite{Krapivsky:2015a}). In the former case, $\eta\r(0)$ is given \textit{a priori}, and thus $\delta \eta\r(0)=0$ by construction. In the latter case, the action should be additionally endowed with some free-energy functional $F[\eta\r(0)]$ that describes the fluctuations of $\eta\r(0)$ in the initial state
\rev{(see e.g.~\cref{eq:F_comb_MFT}).}
Since no such term appears in the action given in \cref{eq:action1,eq:action2}, we interpret our average as quenched, and the term proportional to $\delta \eta\r(0)$ thus gives no 
information. 
\rev{Note that this choice of the initial condition entails no major loss of generality in the calculation of \cref{sec:SP}, because (at odds with SEP in $D=1$ or on the comb) a stationary state is eventually attained by the system, in spite of the infinite-lattice geometry.}

Next, before computing the variation of the third term in \cref{eq:action2}, it is useful to rewrite it as
\begin{align}
    &\sum\r \sum_{\vec \nu} \left[\eta\r \left(1-\eta\rpn \right)\left(e^{- \vec \nu\cdot \vec \varphi\r}-1\right)+\eta\rpn \left(1-\eta\r\right)\left(e^{ \vec\nu\cdot \vec\varphi\r}-1\right)\right] \n\\
    &= \sum_{\vec r,\vec s} \sum_{\vec \nu} \delta_{\vec s,\vec r+\vec \nu} \left[\eta\r \left(1-\eta\s \right)\left(e^{- \vec \nu\cdot \vec \varphi\r}-1\right)+\eta\s \left(1-\eta\r\right)\left(e^{ \vec\nu\cdot \vec\varphi\r}-1\right)\right] ,
    \label{appeq:third}
\end{align}
where we omitted the $t$-dependence so as not to clutter the notation.
As the system is infinite, the boundaries of these sums pose no particular complications, and varying \cref{appeq:third} renders after some algebra
\begin{align}
    &\delta\sum_{\vec r,\vec s} \sum_{\vec \nu} \delta_{\vec s,\vec r+\vec \nu} \left[\eta\r \left(1-\eta\s \right)\left(e^{- \vec \nu\cdot \vec \varphi\r}-1\right)+\eta\s \left(1-\eta\r\right)\left(e^{ \vec\nu\cdot \vec\varphi\r}-1\right)\right] \n\\
    &= \sum\r \sum_{\vec \nu} \left[
        \left(1-\eta\rpn \right)\left(e^{- \vec \nu\cdot \vec \varphi\r}-1\right)-\eta\rpn \left(e^{ \vec\nu\cdot \vec\varphi\r}-1\right)
        +\left(1-\eta\rmn \right)\left(e^{\vec \nu\cdot \vec \varphi\rmn}-1\right)-\eta\rmn \left(e^{ -\vec\nu\cdot \vec\varphi\rmn}-1\right)
        \right] \delta\eta\r \n\\
        &\quad - \sum\r \sum_{\vec \nu} \left[ \eta\r(1-\eta\rpn)e^{-\vec \nu \cdot \vec \varphi\r} -\eta\rpn(1-\eta\r)e^{\vec\nu\cdot \vec \varphi\r} 
        \right]\vec \nu \cdot \delta \vec \varphi\r.
    \label{appeq:3}
\end{align}
Finally, from \cref{eq:genf} we get
\begin{equation}
    \delta \left( 
    \vec \lambda
    \cdot \int_0^T \dd{t} \vec j_{\vec 0}(t) \right) = \vec \lambda
    \cdot \int_0^T \dd{t} \delta\vec j_{\vec 0}(t).
    \label{appeq:4}
\end{equation}
Since the individual variations given in \cref{appeq:1,appeq:2,appeq:3,appeq:4} must vanish independently, collecting their prefactors and equating them to zero finally renders the saddle-point equations reported in \cref{sec:SP}, see \cref{eq:eta,eq:j,eq:theta,eq:varphi}.

\section{Current fluctuations from the microscopic action in 1D}
\label{app:micro_1D}

Since exact results are available for finite systems in 1D~\cite{Derrida:2004}, here we use this setting as a benchmark for the approach we used in Section~\ref{sec:FlucJ2D} for the 2D SEP.

We consider a finite 1D system of $L$ sites, with occupations $\{ \eta_1(t) , \ldots, \eta_L(t) \}$, between two reservoirs. The reservoir on the left can inject particles on site $1$ with rate $\alpha$ and absorb them with rate $\beta$, while the reservoir on the right can inject particles on site $L$ with rate $\gamma$ and absorb them with rate $\delta$. We denote $j_r(t)$ the current from site $r$ to $r+1$ at time $t$. We use the convention that $j_0$ is the current from the left reservoir, and $j_L$ the current towards the right reservoir.

\subsection{Known results in finite 1D systems}

We consider the integrated current $Q_T$ corresponding to the number of particles exchanged between the left reservoir and the system up to time $T$,
\begin{equation}
    \label{eq:DefQTMicroAction}
    Q_T = \int_0^T j_0(t) \dd t
    \:.
\end{equation}
Note that we choose to consider the current injected to site $0$, but we could have equivalently chosen the current through any bond in the system, since in the long-time limit they will have the same leading behaviour~\cite{Derrida:2004}.

Using microscopic calculations, the mean and average of $Q_T$ have been computed in~\cite{Derrida:2004} as
\begin{equation}
    \label{eq:MeanCurrent1Dfinite}
    \moy{Q_T} = \frac{\rho_{\mathrm{L}} - \rho_{\mathrm{R}}}{L+a+b-1} T
    \:,
\end{equation}
\begin{equation}
    \label{eq:FlucCurrent1Dfinite}
    \frac{\moy{Q_T^2} - \moy{Q_T}^2}{T}
    =
    \frac{\rho_{\mathrm{L}} + \rho_{\mathrm{R}} - 2 \rho_{\mathrm{L}} \rho_{\mathrm{R}}}{N}
    + \frac{a(a-1)(2a-1) + b(b-1)(2b-1) - N(N-1)(2N-1)}{3 N^3(N-1)}
    (\rho_{\mathrm{L}} - \rho_{\mathrm{R}})^2
    \:,
\end{equation}
where
\begin{equation}
\label{eq:DefsDerrida}
  a = \frac{1}{\alpha + \beta}
  \:,
  \quad
  b = \frac{1}{\gamma + \delta}
  \:,
  \quad
  \rho_L = \frac{\alpha}{\alpha + \beta}
  \:,
  \quad
  \rho_R = \frac{\gamma}{\gamma + \delta}
  \:,
  \quad
  N = L+a+b-1
  \:.
\end{equation}
Our goal will be to reproduce these results using a microscopic action formalism, as the one used in Section~\ref{sec:FlucJ2D}.

The average occupations and their correlations have also been computed, and read~\cite{Derrida:2004}
\begin{equation}
    \label{eq:MeanOccup1Dfinite}
    \moy{\eta_i} = \frac{\rho_{\mathrm{L}}(L+b-i) + \rho_{\mathrm{R}}(i-1+a)}{L+a+b-1}
    \:,
\end{equation}
\begin{equation}
    \label{eq:CorrelOccup1Dfinite}
    \moy{\eta_i \eta_j} - \moy{\eta_i} \moy{\eta_j}
    = - (\rho_{\mathrm{L}} - \rho_{\mathrm{R}})^2
    \frac{(a+i-1)(b+L-j)}{(L+a+b-1)^2 (L+a+b-2)}
    \:,
    \quad
    \text{for } i < j
    \:.
\end{equation}
Remarkably, combining the two expressions above, the two point correlations can be written in terms of the means in a simple form,
\begin{equation}
    \label{eq:TwoPtFct1D}
  \moy{\eta_i \eta_j} = \moy{\eta_i} \moy{\eta_j}
  + \frac{(\moy{\eta_i} - \rho_{\mathrm{L}})(\moy{\eta_j} - \rho_{\mathrm{R}})}{L + a + b -2}
  \:,
  \quad \text{for }  i< j
  \:.
\end{equation}
This expression will be useful below.

\subsection{Microscopic action formalism}

The formalism of Section~\ref{sec:FlucJ2D} can also be applied to describe this situation, similarly to what was done in~\cite{Lefevre:2007,Saha:2023}. As before, we can write for each site $1 \leq r \leq L$ a microscopic conservation equation,
\begin{equation}
    \eta_{r}(t+\dd t) - \eta_r(t) = (j_{r-1}(t) - j_r(t))\dd t
    \:,
\end{equation}
with the convention that the current $j_r$ counts positively particles leaving site $r$ for site $r+1$. The stochastic dynamics is fully encoded in the currents, which in the bulk take the form,
\begin{equation}
    j_r(t) \dd t = \eta_r(1-\eta_{r+1}) \xi_{r,+}(t) - \eta_{r+1}(1-\eta_r) \xi_{r+1,-}(t)
    \:,
    \quad
    1 \leq r \leq L-1
    \:,
\end{equation}
with $\xi_{r,\pm}$ being random variables, with distribution
\begin{equation}
    \xi_{r,\pm}(t) =  \left\lbrace
    \begin{array}{ll}
        1 & \text{with probability } \dd t \:,  \\
        0 & \text{with probability } 1-\dd t \:.
    \end{array}
    \right.
\end{equation}
These are completed by boundary equations implementing the exchanges with the reservoirs,
\begin{equation}
    j_0(t) \dd t = (1-\eta_1) \zeta_{0,i} - \eta_1 \zeta_{0,a}
    \:,
    \quad
    j_L(t) \dd t = \eta_L \zeta_{L,a} - (1-\eta_L) \zeta_{L,i}
    \:,
\end{equation}
where the random variables $\zeta$ describe the injection or absorption of particles, and are distributed according to
\begin{equation}
    \zeta_{0,i}(t) =  \left\lbrace
    \begin{array}{ll}
        1 & \text{prob. } \alpha \dd t  \\
        0 & \text{prob. } 1- \alpha \dd t
    \end{array}
    \right.
    \:,
    \quad
    \zeta_{0,a}(t) =  \left\lbrace
    \begin{array}{ll}
        1 & \text{prob. } \beta \dd t  \\
        0 & \text{prob. } 1- \beta \dd t
    \end{array}
    \right.
    \:,
\end{equation}
\begin{equation}
    \zeta_{L,i}(t) =  \left\lbrace
    \begin{array}{ll}
        1 & \text{prob. } \gamma \dd t  \\
        0 & \text{prob. } 1- \gamma \dd t
    \end{array}
    \right.
    \:,
    \quad
    \zeta_{L,a}(t) =  \left\lbrace
    \begin{array}{ll}
        1 & \text{prob. } \delta \dd t  \\
        0 & \text{prob. } 1- \delta \dd t
    \end{array}
    \right.
    \:.
\end{equation}
Following the same approach as in Section~\ref{sec:FlucJ2D}, we can write the probability to observe a trajectory $\{ \eta_r(t), j_r(t) \}$ of the occupations and currents up to a time $T$ as
\begin{multline}
  \label{eq:DiscrAction1D}
  P[\eta,j] =
  \int \prod_{r,t} \frac{\dd \varphi_r(t)}{(2\pi)^D}
  \frac{\dd \theta_{r}(t)}{2\pi}
  \exp \Bigg\lbrace
  \int_0^T \dd t \Bigg[
  \I \sum_{r=0}^L \varphi_r j_{r}
  - \sum_{r=1}^L
  \I \theta_r \left( \dt{}{t} \eta_{r}
    - \left(j_{r-1} -j_{r} \right) 
  \right)
  \\
  + \sum_{r=1}^{L-1}
  \Bigg( \eta_{r}(1-\eta_{r+1}) (\e^{- \I \varphi_r}-1)
  + \eta_{r+1}(1-\eta_{r}) (\e^{\I \varphi_r} - 1) \Bigg)
  \\
  + \alpha (1-\eta_1) (\e^{-\I \varphi_0} - 1) + \beta \eta_1 (\e^{\I \varphi_0}-1)
   + \gamma (1-\eta_L) (\e^{\I \varphi_L}-1) + \delta \eta_L (\e^{-\I \varphi_L}-1)
  \Bigg]
  \Bigg\rbrace
  \:.
\end{multline}
The moment generating function of $Q_T$~\eqref{eq:DefQTMicroAction} therefore takes the form,
\begin{multline}
  \label{eq:CGFDiscrAction1D}
  \moy{\e^{\lambda Q_T}}=
  \int \prod_{r,t} \dd \eta_{r}(t) \dd j_r(t)  \frac{\dd \varphi_r(t)}{(2\pi)^D}
  \frac{\dd \theta_{r}(t)}{2\pi}
  \exp \Bigg\lbrace
  \int_0^T \dd t \Bigg[
  \I \sum_{r=0}^L \varphi_r j_{r}
  - \sum_{r=1}^L
  \I \theta_r \left( \dt{}{t} \eta_{r}
    - \left(j_{r-1} -j_{r} \right) 
  \right)
  \\
  + \sum_{r=1}^{L-1}
  \Bigg( \eta_{r}(1-\eta_{r+1}) (\e^{- \I \varphi_r}-1)
  + \eta_{r+1}(1-\eta_{r}) (\e^{\I \varphi_r} - 1) \Bigg)
  \\
  + \alpha (1-\eta_1) (\e^{-\I \varphi_0} - 1) + \beta \eta_1 (\e^{\I \varphi_0}-1)
   + \gamma (1-\eta_L) (\e^{\I \varphi_L}-1) + \delta \eta_L (\e^{-\I \varphi_L}-1)
   + \lambda j_0
  \Bigg]
  \Bigg\rbrace
  \:.
\end{multline}
In this finite 1D system, the occupations and currents are known to be stationary~\cite{Derrida:2004,Derrida:2019,Derrida:2019b}, hence we would like to write an action in terms of their time averaged values only,
\begin{equation}
    \label{eq:AvgValuesDiscrFields}
    \bar\eta_r \equiv \frac{1}{T} \int_0^T \eta_r(t) \dd t
    \:,
    \quad
    \bar{j}_r \equiv \frac{1}{T} \int_0^T j_r(t) \dd t
    \:,
\end{equation}
and similarly for $\theta_r$ and $\varphi_r$. Note that we did not follow the same approach in \cref{sec:SP}, where we kept instead the time dependence of the fields, and found that the saddle-point solutions relax to time-independent values, which must correspond to~\eqref{eq:AvgValuesDiscrFields}. Replacing the fields with their time averages would allow us to write the moment generating function of $Q_T$ as
\begin{equation}
    \label{eq:MomGenFctGuessAction}
  \moy{\e^{\lambda Q_T}} =
  \int \prod_r \dd \bar\eta_r \dd \bar{j}_r  
  \dd \bar\varphi_r
  \dd \bar\theta_{r}
  \exp \left[ T \: \mathcal{S}[\{ \bar\eta_r, \bar{j}_r, \bar\varphi_r, \bar\theta_r\}] \right]
  \:,
\end{equation}
with an effective action $\mathcal{S}$ to be determined. Such a representation would be extremely convenient, since for large $T$ the integral can be computed by a simple saddle point, and thus yield the cumulant generating function of $Q_T$ as
\begin{equation}
    \label{eq:RelJCumul}
    \ln \moy{\e^{\lambda Q_T}} \underset{T \to \infty}{\simeq}
    T \: \mathcal{S}[\{ \bar\eta_r^\star, \bar{j}_r^\star, \bar\varphi_r^\star, \bar\theta_r^\star\}]
    \:,
\end{equation}
where we have denoted $\{ \bar\eta_r^\star, \bar{j}_r^\star, \bar\varphi_r^\star, \bar\theta_r^\star\}$ the saddle point. In particular, this implies that the cumulants of $Q_T$ can be easily computed as
\begin{equation}
    \dt{}{\lambda} \ln \moy{\e^{\lambda Q_T}} \underset{T \to \infty}{\simeq}
    T \: \bar{j}_0^\star
    \:.
\end{equation}
In the following, we will drop the $\star$ to lighten the notations. The main remaining difficulty is to obtain the effective action $\mathcal{S}$.

\begin{figure}
    \centering
    \includegraphics[height=0.3\textwidth]{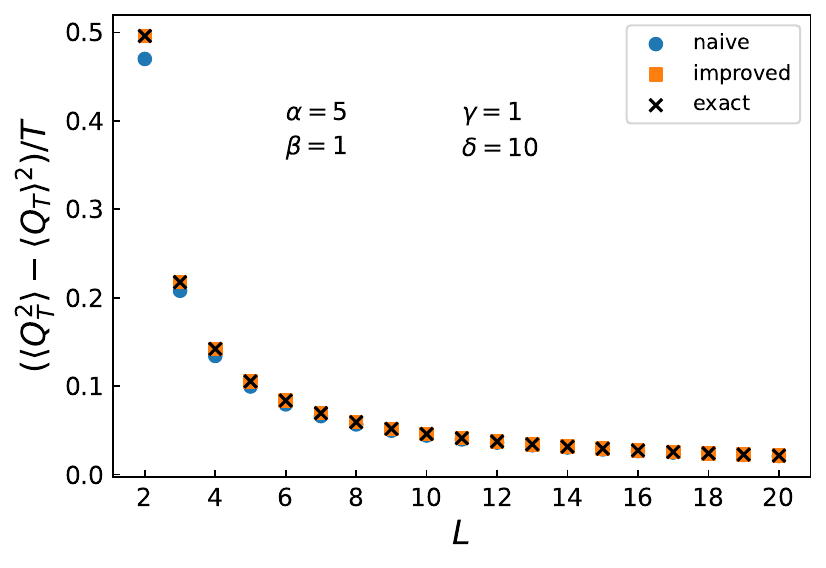}
    \includegraphics[height=0.3\textwidth]{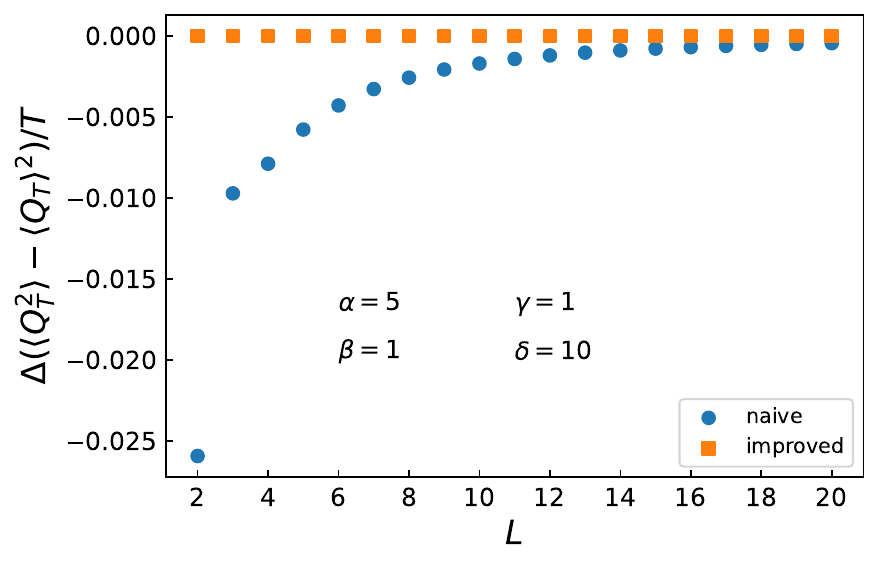}
    \caption{Left: predictions for the fluctuations of the current $\frac{\moy{Q_T^2} - \moy{Q_T}^2}{T}$, obtained numerically by optimising the naive action~\eqref{eq:DiscrAction1DNaive} and the improved action~\eqref{eq:DiscrAction1DImproved}, compared to the exact value~\eqref{eq:FlucCurrent1Dfinite}, for different system sizes. Right: difference between the values of the fluctuations of $Q_T$ obtained from the actions~\eqref{eq:DiscrAction1DNaive} and~\eqref{eq:DiscrAction1DImproved} and the exact one~\eqref{eq:FlucCurrent1Dfinite}. The plots are for $\alpha = 5$, $\beta = 1$, $\gamma = 1$ and $\delta = 10$.}
    \label{fig:CompFluc1Dfinite}
\end{figure}

If we naively replace $\eta_r$ and $j_r$ in the generating function~\eqref{eq:CGFDiscrAction1D} by their time averages $\bar\eta_r$ and $\bar{j}_r$, and do the same for $\theta_r$ and $\varphi_r$, we get a naive action
\begin{multline}
  \label{eq:DiscrAction1DNaive}
  \mathcal{S}_{\mathrm{naive}}[\{ \bar\eta_r, \bar{j}_r, \bar\varphi_r, \bar\theta_r\}]
  = 
  \I \sum_{r=0}^L \bar\varphi_r \bar{j}_{r}
  + \sum_{r=1}^L
  \I \bar\theta_r \left(
     \bar{j}_{r-1} - \bar{j}_{r}
  \right)
  + \sum_{r=1}^{L-1}
  \Bigg( \bar\eta_{r}(1-\bar\eta_{r+1}) (\e^{- \I \bar\varphi_r}-1)
  + \bar\eta_{r+1}(1-\bar\eta_{r}) (\e^{\I \bar\varphi_r} - 1) \Bigg)
  \\
  + \alpha (1-\bar\eta_1) (\e^{-\I \bar\varphi_0} - 1) + \beta \bar\eta_1 (\e^{\I \bar\varphi_0}-1)
   + \gamma (1-\bar\eta_L) (\e^{\I \bar\varphi_L}-1) + \delta \bar\eta_L (\e^{-\I \varphi_L}-1)
   + \lambda \bar{j}_0
  \:.
\end{multline}
In this form, we can optimise this effective action to obtain equations satisfied by $\{ \bar\eta_r, \bar{j}_r, \bar\varphi_r, \bar\theta_r\}$. These equations are rather cumbersome and cannot be solved analytically, so we do not reproduce them here. However, for $\lambda = 0$ they can be solved, and give $\bar\varphi_r = 0$, $\bar\theta_r = 0$, $\bar\eta_r = \moy{\eta_r}$ given by~\eqref{eq:MeanOccup1Dfinite}, and $\bar{j}_r = \moy{Q_T}/T$ given by~\eqref{eq:MeanCurrent1Dfinite}. This shows that the naive action in~\eqref{eq:DiscrAction1DNaive} properly reproduces the known results for the average quantities. To investigate whether this is also true for the fluctuations, we expand the optimisation equations at first order in $\lambda$, using
\begin{equation}
    \bar\eta_r = \bar\eta_r^{(0)} + \lambda \bar\eta_r^{(1)} + O(\lambda^2)
    \:,
    \quad
    \bar{j}_r = \bar{j}_r^{(0)} + \lambda \bar{j}_r^{(1)} + O(\lambda^2)
    \:,
    \quad
    \bar\varphi_r = \lambda \bar\varphi_r^{(1)} + O(\lambda^2)
    \:,
    \quad
    \bar\theta_r = \lambda \bar\theta_r^{(1)} + O(\lambda^2)
    \:,
\end{equation}
and solve the equations at first order in $\lambda$. In light of~\eqref{eq:RelJCumul}, $\bar{j}_0^{(1)}$ is the variance of $Q_T$, and should thus reproduce~\eqref{eq:FlucCurrent1Dfinite}. Since the equations are cumbersome, we implement this procedure with \texttt{Mathematica} and solve them numerically for different system sizes $L$. The result is shown in Fig.~\ref{fig:CompFluc1Dfinite}. We observe a discrepancy between the values predicted from this naive action and the exact result~\eqref{eq:FlucCurrent1Dfinite}. One possible explanation is that, when replacing the value $\eta_r(t)$ by its temporal average $\bar\eta_r$, we have lost track of the correlations between the sites, which are important since the original action~\eqref{eq:CGFDiscrAction1D} involves products $\eta_r(t) \eta_{r+1}(t)$, whose temporal average is not $\bar\eta_r \bar\eta_{r+1}$.

To improve the naive action~\eqref{eq:DiscrAction1DNaive}, we can try to add ``by hand'' the missing two-point correlations between different sites~\eqref{eq:TwoPtFct1D}, i.e.~by making into the moment generating function~\eqref{eq:CGFDiscrAction1D} the substitution
\begin{equation}
    \eta_r(t) \eta_{r+1}(t) \longrightarrow \bar\eta_r \bar\eta_{r+1}
    + \frac{(\bar\eta_r - \rho_L)(\bar\eta_{r+1} - \rho_R)}{L+a+b-2}
    \:,
\end{equation}
with $a$, $b$, $\rho_{\mathrm{L}}$ and $\rho_{\mathrm{R}}$ defined in~\eqref{eq:DefsDerrida}.
This leads us to write an improved guess for the action,
\begin{multline}
  \label{eq:DiscrAction1DImproved}
  \mathcal{S}_{\mathrm{improved}}[\{ \bar\eta_r, \bar{j}_r, \bar\varphi_r, \bar\theta_r\}]
  = 
  \I \sum_{r=0}^L \bar\varphi_r \bar{j}_{r}
  + \sum_{r=1}^L
  \I \bar\theta_r \left(
     \bar{j}_{r-1} - \bar{j}_{r}
  \right)
  + \sum_{r=1}^{L-1}
  \Bigg( \bar\eta_{r}(1-\bar\eta_{r+1}) (\e^{- \I \bar\varphi_r}-1)
  + \bar\eta_{r+1}(1-\bar\eta_{r}) (\e^{\I \bar\varphi_r} - 1) \Bigg)
  \\
  + \sum_{r=1}^{L-1} 2( 1-\cos \bar\varphi_r ) \frac{(\bar\eta_r - \rho_L)(\bar\eta_{r+1} - \rho_R)}{L+a+b-2}
  \\
  + \alpha (1-\bar\eta_1) (\e^{-\I \bar\varphi_0} - 1) + \beta \bar\eta_1 (\e^{\I \bar\varphi_0}-1)
   + \gamma (1-\bar\eta_L) (\e^{\I \bar\varphi_L}-1) + \delta \bar\eta_L (\e^{-\I \varphi_L}-1)
   + \lambda \bar{j}_0
  \:.
\end{multline}
Following the same procedure as for $\mathcal{S}_{\mathrm{naive}}$, we check that this action still gives the correct average values for the currents~\eqref{eq:MeanCurrent1Dfinite} and occupations~\eqref{eq:MeanOccup1Dfinite}. In addition, we compute numerically the current fluctuations from this new action, which is again shown in Fig.~\ref{fig:CompFluc1Dfinite}. Remarkably, the new action~\eqref{eq:DiscrAction1DImproved} properly reproduces the fluctuations of $Q_T$ for any system size. We expect that this feature will hold in any dimension. Since for infinite systems at equilibrium the correlations between two sites vanish (see~\cite{Derrida:2004} or \cref{equilibrium_distribution} above for the infinite comb), the naive action~\eqref{eq:DiscrAction1DNaive} (and its generalisation to higher dimensional systems) 
is expected to yield the correct fluctuations of $Q_t$
for infinite systems at equilibrium.


%

\end{document}